%% file: soar_tess_iii.tex
\documentclass[twocolumn]{aastex63}
\usepackage{amsmath}
\makeatletter
\renewcommand{\tablecomments}[1]{\vskip1pt{\small\vskip1sp\indent
\vrule height 11pt depth 2pt width 0pt\currtabletypesize
{\sc Note}. {#1}\vskip1pt}}
\renewcommand{\subsubsection}{\@startsection{subsubsection}{3}{\z@}%
  {1.5ex plus .5ex minus .2ex}{.25ex plus .05ex}%
  {\small\itshape\center}}
\makeatother
\newcommand{\tess}{\textit{TESS}}

\newcommand{\rearth}{\ensuremath{R_\oplus}}
\newcommand{\msun}{\ensuremath{M_\odot}}
\shorttitle{Mapping the Stellar Companion Deficit}
\shortauthors{Ziegler et al.}

\begin{document}

\title{SOAR TESS Survey III: Mapping the Stellar Companion Deficit around TESS Planet Candidates}

\correspondingauthor{Carl Ziegler}
\email{Carl.Ziegler@sfasu.edu}

\author[0000-0002-0619-7639]{Carl Ziegler}
\affiliation{Department of Physics, Engineering and Astronomy, Stephen F. Austin State University, Nacogdoches, TX 75962, USA}
\author[0000-0002-2084-0782]{Andrei Tokovinin}
\affiliation{Cerro Tololo Inter-American Observatory/NSF's NOIRLab, Casilla 603, La Serena, Chile}
\author[0000-0001-7124-4094]{C\'esar Brice\~no}
\affiliation{Cerro Tololo Inter-American Observatory/NSF's NOIRLab, Casilla 603, La Serena, Chile}
\author[0000-0001-9380-6457]{Nicholas M. Law}
\affiliation{Department of Physics and Astronomy, The University of North Carolina at Chapel Hill, Chapel Hill, NC 27599-3255, USA}
\author[0000-0003-3654-1602]{Andrew W. Mann}
\affiliation{Department of Physics and Astronomy, The University of North Carolina at Chapel Hill, Chapel Hill, NC 27599-3255, USA}
\author[0009-0004-2552-9581]{Seth Batson}
\affiliation{Department of Physics, Engineering and Astronomy, Stephen F. Austin State University, Nacogdoches, TX 75962, USA}

\begin{abstract}
We present the cumulative SOAR TESS Survey, based on HRCam speckle observations of 2,982 targets, and report 433 companion measurements obtained after Survey II. In a demographic sample of 1,199 dwarf primaries, we compare detected companions with secondary-to-primary mass ratios $q=M_2/M_1\geq0.4$ to a field-star population after accounting for each observation's detection limits. We find 24 companions within 50 au versus 84.8 expected and 48 within 100 au versus 124.4 expected. The companion frequencies relative to the field expectation are $S_{50\,{\rm au}}=\allowbreak0.291^{+0.062}_{-0.054}$ and $S_{100\,{\rm au}}=\allowbreak0.391^{+0.058}_{-0.053}$. A continuous model in which the deficit weakens with separation reaches the midpoint between close-binary suppression and the field rate at $a_{50}=74$ au (57--101 au). The Gaia-matched subset contains 88 companions versus 69.6 expected, a difference consistent with the model once uncertainties in the field comparison and Gaia selection are included. A model that preserves the total number of binaries but shifts the field distribution toward 100 au fits poorly. Under the limiting assumption that all retained planets orbit the catalog primary, the fitted law implies reductions of 17.0\% in volume-limited planet yield over $q\geq0.4$ and 23.3\% when extended to all mass ratios. M-dwarf hosts show a similar close-companion deficit, but the smaller sample prevents a precise comparison with solar-type hosts.
\end{abstract}


\section{Introduction}\label{sec:intro}

The Transiting Exoplanet Survey Satellite (\tess; \citealt{Ricker2015}) searches bright, nearby stars for transiting planets. A neighboring star can dilute a transit, bias the inferred planet radius, reduce the detectability of small planets, or produce the signal through an eclipsing binary. High-angular-resolution imaging is therefore an important part of planet validation and characterization \citep{Furlan2017,Matson2019,Howell2021Program}.

Resolved sources can test planet formation in binaries once physical association is established or treated probabilistically. A deficit of stellar companions to exoplanet hosts below approximately 100 au is well documented \citep{Kraus2016,Ziegler2020,Ziegler2021,Howell2021,Lester2021,Clark2022,Littlefield2024}. The reference is the companion frequency of similar-mass field stars, and we quantify the deficit with a suppression factor $S=f_{\rm TOI}/f_{\rm field}$. Thus $S=1$ means the TOI and field companion frequencies agree, whereas $S<1$ means that companions are underrepresented among TOI hosts. The rise of $S$ toward unity at larger separation is called ``recovery'' throughout this paper. \citet{Kraus2016}, for example, found that planet occurrence in binaries inside a characteristic separation of $47^{+59}_{-23}$ au was $0.34^{+0.14}_{-0.15}$ times that of wider binaries and single stars.

A volume-limited radial-velocity survey found no companions within 50 au of its planet hosts \citep{Hirsch2021}, and a direct Kepler occurrence analysis measured 50--58\% fewer planets in its adopted small-separation binary samples \citep{Sullivan2026}. Unlike our census of already identified TOIs, those studies start from a defined sample of searched stars and model the probability of detecting a planet. They therefore motivate, but do not determine, the physical interpretation of our companion-frequency measurement.

A planet-selected sample has an additional bias in the opposite direction from a close-companion deficit: a binary contains two potential planet-host stars and therefore two opportunities to enter the TOI catalog. For a near-equal pair with comparable planet populations and detectability, the probability that at least one component supplies a detected transit can approach twice the single-star probability. The extra opportunity is much smaller for a faint secondary because its planets are more strongly diluted and small planets may fall below the detection threshold. Consequently, finding fewer close binaries among TOIs despite this two-component advantage makes the underlying deficit in planet occurrence per star stronger than the raw companion-frequency ratio would suggest.

Several mechanisms can produce this deficit. Binary torques truncate circumstellar disks to a fraction of the binary periastron separation \citep{ArtymowiczLubow1994}. Close binaries also lose primordial disks more rapidly than single stars and wide binaries \citep{Cieza2009,Kraus2012}. Truncation reduces the available reservoirs of gas and solids, shortens the disk evolution time, and accelerates radial drift of solids \citep{Zagaria2021,Zagaria2023}. Recent global formation calculations find that truncation reduces the pebble supply and steadily suppresses the formation of planets more massive than Mars below binary separations of about 160 au \citep{Venturini2026}. A companion can also excite destructive planetesimal collisions or destabilize planets after formation \citep{SilsbeeRafikov2021,Marzari2019,MoeKratter2021}. These mechanisms predict a gradual recovery of the planet-host companion frequency toward the field value, rather than a universal cutoff, and their strength should also depend on binary mass ratio and eccentricity.

Selection effects act in competing directions. Magnitude-limited samples overrepresent unresolved binaries, and two potential hosts increase the chance that a binary supplies a planet candidate, but transit dilution reduces planet detectability. Catalog cuts can also remove binaries before imaging; \citet{Littlefield2024} found that requiring complete TIC parameters preferentially excludes some companions at 0.1--1.2 arcsec. At these intermediate angular separations Gaia may not provide two complete source entries, the TIC can contain merged or incomplete stellar parameters, and follow-up teams may deprioritize an ambiguous target. Occurrence calculations that include unresolved companions find larger corrections than analyses that adjust only detected planet radii \citep{Bergsten2026}. We treat the measured contrast curve, target selection, and planet-detection efficiency as separate parts of the calculation.

Gaia complements speckle imaging at wider separations. Common parallax and proper motion identify likely bound companions, and recent Gaia searches have extended the wide-companion census of TESS planet candidates \citep{EelesNolle2025,Mugrauer2025}. Gaia is nevertheless incomplete where HRCam is most sensitive \citep{Gaia2023,GaiaDR3Data,ElBadry2021}. Combined-data tests show why this intermediate regime requires several diagnostics: Gaia imaging, RUWE, and high-resolution imaging overlap but leave different parts of the companion parameter space unconstrained \citep[see Figure 7 and the surrounding discussion in][]{Wood2021}. Unresolved companions can instead inflate Gaia's renormalized unit-weight error (RUWE) or produce excess radial-velocity noise. The \textsc{paired} framework models the latter against stars of similar color and magnitude and is most sensitive to semimajor axes below roughly 10 au \citep{Chance2025}. These diagnostics probe different observables and do not always respond to the same component in a hierarchical system.

Here we analyze the complete SOAR/HRCam program as a cumulative imaging census. We publish the post-Paper-II measurements, document how observing lists were constructed, calculate target-level contrast curves and dilution corrections, perform a Gaia DR3 common-motion search, and compare the mass-ratio-matched close-source distribution with a sensitivity-convolved field-binary model. The primary measurement is the resolved-source frequency among cataloged TESS Objects of Interest (TOIs) relative to that field model. Translating it into intrinsic planet occurrence additionally requires association, reliability, and transit-completeness information. We also examine how the companion frequency depends on separation, planet radius, and stellar mass, and what it may imply for planet occurrence.

\section{Observations and Analysis Sample}\label{sec:data}

The full imaging census and the demographic sample serve different purposes. The former records everything HRCam observed; the latter retains the targets that can be compared consistently with a field population. Using each target's measured sensitivity prevents repeat visits, uneven observing depth, and missing catalog parameters from being mistaken for real differences in companion frequency.

\subsection{Cumulative HRCam observations}

We began with the cumulative SOAR/HRCam observation table, the HRCam contrast-limit file, and a frozen 2026 July 30 table of current TESS planet dispositions and stellar properties \citep{Stassun2019,Guerrero2021,ExoFOPTESS}. The observation table contains 3,118 records, 3,054 independent TIC--date visits, and 2,982 unique TIC identifiers. Multiple records on one date usually describe different components of a triple or higher-order configuration. We count each TIC--date once when describing the observing program.

The observations reported here and in Surveys I and II were obtained with the High-Resolution Camera (HRCam) speckle instrument at the 4.1-m Southern Astrophysical Research telescope in Chile \citep{Ziegler2020,Ziegler2021,Tokovinin2018}. For the TESS survey, HRCam records a $200\times200$-pixel region after $2\times2$ binning, with an effective scale of approximately 30 mas pixel$^{-1}$ and a 6 arcsec field. We used the $I$ filter ($\lambda_{\rm cen}=824$ nm, $\Delta\lambda=170$ nm), which is similar to the effective TESS bandpass. A standard observation consists of two data cubes of 400 short-exposure frames. The pipeline derives the power spectrum and speckle autocorrelation function (ACF); binary and triple-star fringes in the power spectrum yield separation, position angle, and $\Delta I$. Internal calibration repeatability is approximately 0.1\% in separation and $0.1^\circ$ in position angle \citep{Tokovinin2018}. For comparisons with external astrometry, we adopt additional uncertainty floors of 3.8 mas in separation and $0.95^\circ$ in position angle, based on the SOAR residuals compiled by \citet{Mann2019}; the typical differential-photometry uncertainty is 0.1 mag. The pipeline also supplies a six-point $5\sigma$ contrast-limit curve. The program's brightness-limited target selection enhances the combined-light Malmquist preference for unresolved binaries. Recent annual releases continue to recover close pairs that do not appear as two Gaia sources \citep{Tokovinin2026}.

Table~\ref{tab:summary} summarizes the cumulative survey accounting used throughout this paper.

Throughout this paper, ``cumulative survey'' refers to all SOAR/HRCam observations in the frozen archive, ``post-Paper-II measurements'' refers only to the 433 measurements obtained after 2020 December 31 and published in Appendix~\ref{app:newcompanions}, and ``demographic sample'' refers to the restricted 1,199-host dwarf-primary subset defined below. These samples overlap, but their counts are not interchangeable.

For targets observed more than once, the demographic calculations retain unique detections at the latest epoch. In the component labels, A is the catalog primary; A--B and A--C identify companions measured relative to A; A--BC identifies A relative to an unresolved BC subsystem; and B--C or Ba--Bb identify an inner pair in the B subsystem. The survey notes also use shorthand such as ``BC'' and ``triple''; all of these forms are included when constructing the source catalog. Every retained post-Survey-II source was required to have a SOAR/HRCam ACF product; earlier sources were checked against the published Surveys I and II measurement catalogs. An imaging note by itself was not accepted as evidence of a SOAR detection. The cumulative latest-epoch survey catalog contains 731 resolved-source detections around 689 hosts, including 472 hosts with a source inside 1.5 arcsec and 631 inside 3 arcsec. These are cumulative survey yields, not occurrence rates, because the program includes validation targets, false positives, and repeated observations.

Throughout the paper, a ``resolved source'' is an observational detection. We use ``candidate visual companion'' when physical association is unknown, ``common-motion companion'' for a Gaia-selected pair, and ``binary'' only for a simulated field system or a pair with evidence of association. A SOAR source remains of unknown association until common parallax and proper motion, multi-epoch relative astrometry, or other evidence is stated explicitly. Apparent projected separation is the angular separation multiplied by the primary's distance; for an unassociated source it serves only as a convenient scale.

\begin{deluxetable}{lr}
\tablecaption{Cumulative SOAR/HRCam survey summary\label{tab:summary}}
\tablehead{\colhead{Quantity}&\colhead{Value}}
\startdata
Observation records & 3,118\\
Independent TIC--date visits & 3,054\\
Unique TIC identifiers & 2,982\\
Visits with reliable six-point curves & 2,728\\
Targets with a selected contrast curve & 2,684\\
Latest-epoch component detections & 731\\
Hosts with any resolved companion & 689\\
Retained signal entries (hosts) & 2,351 (2,197)\\
Confirmed planet entries (hosts) & 377 (312)\\
Close-binary analysis hosts & 1,412\\
Mass-ratio-matched demographic hosts & 1,199\\
Empirical $q\geq0.4$ visual companions & 162\\
\enddata
\tablecomments{Planet counts refer to the frozen disposition table and will change as candidates are confirmed or rejected. The confirmed-planet count describes the surveyed sample regardless of whether the SOAR observation contributed to an individual confirmation. The principal survey products are the 731-component imaging census and its target-level contrast-limit model; the raw companion fraction is descriptive.}
\end{deluxetable}

\subsection{Observing-list construction}\label{sec:targetselection}

Any use of known multiplicity in assembling the observing list could create an artificial companion deficit. We therefore reconstruct how new targets entered the program and distinguish that process from repeat astrometry of systems already counted in the survey.

New SOAR targets were selected from the TOI catalog without using multiplicity information. TICs already observed with SOAR speckle imaging were excluded from the new-target list for each run. The remaining TOIs were filtered by their current disposition and survey status, with known planets and systems already extensively characterized by surveys such as WASP and HAT removed from the ordinary queue. Targets were then prioritized by suitability for HRCam, principally TESS-band magnitude $T\lesssim13$, consistency between the TIC target and the nearby Gaia source, and observability from SOAR during the night. No RUWE, Gaia neighbor count, visual-companion flag, or other indicator of multiplicity entered this new-target selection.

Previously observed binaries were maintained separately at lower observing priority for repeat astrometry and were used to fill gaps in the primary observing sequence. Because these systems had already entered the survey, their binary status could affect whether an additional epoch was obtained but could not add a new target to the demographic denominator. Individual runs necessarily differed in date, sky accessibility, and occasionally in limiting magnitude, and candidate false positives were sometimes observed for validation. This distinction remained the same throughout the program: ordinary new targets were chosen by TOI status, brightness, sky position, and observability, while multiplicity information was used only to prioritize repeat observations of systems already in the survey. The demographic analysis counts each TIC once and applies target-specific sensitivity, so repeated observations of known binaries do not increase the number of companion hosts in the parent sample.

\subsection{Target-level contrast curves}\label{sec:curves}

A nondetection carries weight only over the contrasts and separations probed by that observation. The measured curve for each target determines which synthetic field companions HRCam would have recovered, allowing observing depth to vary realistically across the sample.

The survey reduction provides reliable six-point contrast limits for 2,728 visits and 2,684 unique targets. We treat these limits as direct HRCam survey products. Each record gives the measured angular working range and corresponding $5\sigma$ contrasts; no contrast-limit point is inferred from an external companion catalog.

Repeated observations were reduced to one physical curve per target. We selected the visit with the largest mean contrast at 0.1, 0.2, 1, and 3 arcsec. We did not combine the best point from different epochs, which would create a curve that was never measured in one observation. The occurrence model uses these target-level curves.

\begin{figure}[t]
\centering
\includegraphics[width=\columnwidth]{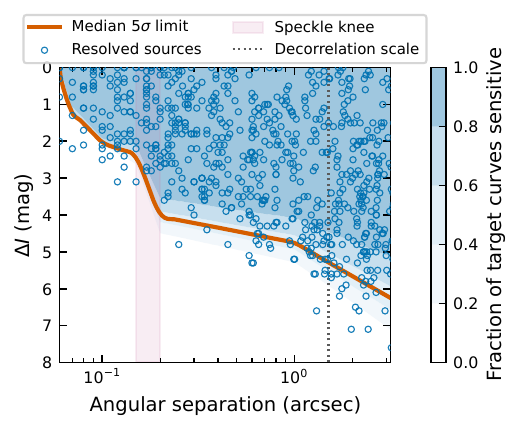}
\caption{Fraction of target-level HRCam curves sensitive to a source at each angular separation and $I$-band contrast. Light-blue bands show completeness in 20\% intervals, the orange line is the smoothed median $5\sigma$ contrast limit, and open circles are resolved sources. The shaded 0.15--0.2 arcsec interval marks the ``speckle knee,'' caused principally by instrumental power-spectrum distortions around the central ACF peak \citep{Tokovinin2010}; the dotted line marks the approximately 1.5 arcsec scale beyond which speckle decorrelation limits further improvement.}\label{fig:sensitivity}
\end{figure}

Figure~\ref{fig:sensitivity} shows the detection limits. Contrast sensitivity improves rapidly outside the diffraction limit. We call the change in slope near 0.15 arcsec the ``speckle knee.'' It is caused mainly by spurious structure around the central ACF peak from instrumental distortions of the power spectrum, not by the 40 mas width of the peak itself; injection--recovery tests show the same transition \citep[Figure 9]{Tokovinin2010}. The median curve then rises more slowly and is nearly flat beyond about 1.5 arcsec, where decorrelation of the speckle pattern limits further improvement.

The corresponding target counts and sensitivity percentiles are listed in Table~\ref{tab:sensitivity}.

\begin{deluxetable}{rrrrr}
\tablecaption{Target-level HRCam contrast limits\label{tab:sensitivity}}
\tablehead{\colhead{$\rho$ (arcsec)}&\colhead{$N$}&\colhead{16th (mag)}&\colhead{Median (mag)}&\colhead{84th (mag)}}
\startdata
0.10 &2600&1.90&2.04&2.10\\
0.20 &2683&3.43&4.04&4.56\\
1.00 &2683&3.95&4.75&5.35\\
1.50 &2684&4.31&5.28&5.99\\
3.00 &2674&4.91&6.19&7.13\\
\enddata
\tablecomments{The last three columns give $\Delta I_{5\sigma}$ contrast limits in magnitudes. Values are interpolated in log separation only within each measured six-point curve. $N$ varies because no extrapolation is made. The median contrast limit rises from 2.04 mag at 0.10 arcsec to 6.19 mag at 3 arcsec.}
\end{deluxetable}

\subsection{Close-binary demographic sample}

The full archive provides the companion census, whereas a frequency comparison requires a more uniform stellar sample. We select dwarf primaries for which observed and simulated companions can share one mass-ratio scale, while retaining blended systems whose incomplete catalog parameters would otherwise make the sample multiplicity dependent.

The close-binary calculation uses a more restricted sample than the survey census. We retained TICs with at least one current confirmed planet (CP), known planet (KP), planet candidate (PC), or ambiguous planet candidate (APC); we refer to those CP/KP/PC/APC entries as ``retained signals.'' We rejected a host if any retained signal exceeded 15~\rearth\ and required $3900<T_{\rm eff}<7400$ K, a known distance, and a matched HRCam curve.

Catalog mass was available for most hosts. For the 94 otherwise eligible targets without one, we interpolated stellar mass from $T_{\rm eff}$ using the empirical dwarf sequence of \citet{PecautMamajek2013}; the adopted values and provenance are released. Omitting these systems would be a strongly nonrandom cut because missing catalog parameters are concentrated among blended, resolved systems. Stellar mass enters the conversion of simulated period to semimajor axis only as $M_\star^{1/3}$, so the population model is less sensitive to a moderate mass error than to excluding the target.

The individual interpolated masses can nevertheless have especially large errors. For the closest pairs, the TIC temperature, radius, and luminosity were derived from photometry that blends both stars; fitting an unresolved binary as one star can substantially bias the inferred stellar parameters \citep{FurlanHowell2020}. The same bias propagates to the catalog planet radius and can weaken the planetary interpretation of a candidate. We retain these systems to avoid a multiplicity-dependent selection cut, but flag their stellar and planet properties as less secure than those based on resolved photometry or spectroscopy.

Disposition labels alone are insufficient at the planet--brown-dwarf boundary. We checked the retained CP, KP, and PC rows against the primary discovery literature and removed 12 transiting brown dwarfs whose labels in the frozen catalog had not all been updated \citep{Anderson2011,Carmichael2020,Carmichael2021,Subjak2020,Persson2019,Grieves2021,Carmichael2022,Psaridi2022}. Seven would otherwise pass the remaining cuts. These requirements define a broad 1,412-host analysis sample.

The primary demographic calculation further requires $0.59\leq M_1/\msun\leq1.75$ and $\log g\geq4.0$, with $\log g$ calculated from the adopted mass and catalog radius. This 1,199-host dwarf-primary sample is the range in which we use the Pecaut--Mamajek dwarf sequence to convert between $I$-band contrast and mass ratio. It is 1.83 times larger than the 655-star culled solar-type sample used for the Paper II demographic analysis \citep{Ziegler2021}. Applying the same mass--luminosity relation to the observed and simulated populations gives both populations a consistent $q\geq0.4$ selection. Sixty-six of these hosts use a temperature-based mass; retaining them is important because they contain 17 of the 24 observed sources inside 50 au and 40 of the 48 inside 100 au.

Within the cumulative 1,199-host dwarf-primary demographic sample, 193 hosts have at least one resolved SOAR source inside 3 arcsec. This 193-host subset is, on average, more distant and more likely to contain a large-radius candidate than the full demographic sample (Appendix~\ref{app:selectionfigures}). These shifts reflect heterogeneous TOI selection and residual false-positive contamination, so the analysis uses target-specific sensitivity rather than a raw resolved-source fraction.

\section{Cumulative Companion Census}\label{sec:companions}

The census establishes where HRCam found nearby sources; the dilution calculation then quantifies how their light biases transit depths and planet radii under an assumed host star. Physical association and identifying which star produces the transit require additional observations, so those questions remain separate from both products.

\subsection{Companions and planet hosts}

We converted angular separation to apparent projected separation using the adopted target distance. For every demographic host, the Pecaut--Mamajek dwarf sequence gives the primary's intrinsic absolute $I$ magnitude from the adopted $M_1$; this is a single-star model magnitude, not the blended catalog $I$ magnitude. Adding the measured $\Delta I$ gives the candidate secondary magnitude and hence an empirical mass ratio. The observed and simulated samples both require $q\geq0.4$. As a robustness check, a fixed optical cut of $\Delta I\leq5.1$ retains 25 rather than 24 sources inside 50 au and 51 rather than 48 inside 100 au. This small change leaves the close deficit intact; all primary results use the empirical mass-ratio cut.

Within the cumulative 1,199-host demographic sample, the mass-ratio and separation cuts retain 162 candidate visual companions between apparent projected separations of 1 and 5,000 au around 158 hosts. Their median apparent projected separation is 214 au. Twenty-four sources lie inside 50 au, and 48 sources around 48 hosts lie inside 100 au. The latest-epoch catalog counts each resolved source once; a repeated archive measurement outside 300 au is therefore excluded from the source count, without affecting the close-source results. For a physically associated pair, the plotted value is the instantaneous sky-plane separation rather than a semimajor axis; for an unassociated source it is an angular-distance scaling rather than a physical binary separation.

For every post-Paper-II component measurement referenced to A and having the required Gaia information, we estimate the chance-alignment probability reported in the post-Paper-II measurement catalog. We follow the local-surface-density method commonly used in high-angular-resolution multiplicity surveys \citep[e.g.,][]{Correia2006,Woellert2015}. Where the available target-centered Gaia DR3 cone extends beyond 10 arcsec, we use the comparison region from 10 arcsec to the smaller of 10,000 au and 600 arcsec. We count sources brighter than the candidate's estimated $G$ magnitude, using $\Delta I\simeq\Delta G$ and capping the limiting magnitude at $G=20.7$. If their local surface density is $\Sigma$, the Poisson plug-in probability of at least one such background source within the measured separation is
\begin{equation}
P_{\rm bg}=1-\exp(-\pi\rho^2\Sigma).
\label{eq:pbg}
\end{equation}
This target-specific diagnostic allows for the strong dependence of background density on Galactic position and companion brightness. Measurements for which the available comparison cone does not reach 10 arcsec are left without a value rather than assigned a probability from an undersized annulus. The catalog distances are retained as adopted values; for very distant or unassociated sources, the tabulated projected separation is only an angular-distance scaling. Of the 393 measurements for which $P_{\rm bg}$ can be calculated, two (0.5\%) have $P_{\rm bg}>0.1$; both also exceed 0.2, and the largest value is 0.365. The remaining 40 of the 433 measurements lack a calculated value because they are inner-subsystem measurements, required Gaia information is unavailable, or the comparison cone does not extend beyond 10 arcsec. A tabulated zero means that no comparison source met the local cut; it is a plug-in estimate, not proof of association.

The frozen planet table contains 2,351 retained CP, KP, PC, or APC entries around 2,197 surveyed TICs, including 377 confirmed-planet entries around 312 TICs. Appendix~\ref{app:selectionfigures} shows their descriptive period--radius distribution and identifies the catalog window-function artifact near 700--750 days. The demographic analysis uses the 162-source mass-ratio-matched sample, not that descriptive planet distribution.

\subsection{System dilution corrections}\label{sec:radius}

Extra light makes a transit shallower and the planet appear smaller. We provide a uniform first-order correction for a transit of the catalog primary; a final radius additionally requires deblended stellar properties and identification of the transited component.

If a planet transits the primary star and all resolved companion light enters the photometric aperture, the system correction is
\begin{equation}
X_{R,A}=\left(1+\sum_j10^{-0.4\Delta I_j}\right)^{1/2}.
\label{eq:primarycorr}
\end{equation}
Here $X$ denotes the multiplicative planet-radius correction, the subscript $R$ denotes radius, and $A$ specifies the assumption that the catalog primary is transited. We sum the flux ratios of all latest-epoch companions inside 3 arcsec. The $I$ filter is a useful but imperfect proxy for the TESS bandpass. These uniform first estimates can be refined with resolved colors and target-specific aperture modeling.

The 631 cumulative-survey hosts with at least one latest-epoch SOAR source inside 3 arcsec have a median correction of 1.070 and a 90th percentile of 1.326. Corrections exceed 1.05 for 357 systems, 1.10 for 254, 1.20 for 140, and 1.30 for 86. The maximum is 1.456 in a hierarchical system. At fixed planet mass, the corrected bulk density is multiplied by $X_{R,A}^{-3}$; the median and 90th-percentile radius corrections therefore lower the inferred density by 18\% and 57\%, respectively. For small planets, even a modest radius revision can change whether a planet is interpreted as a predominantly rocky super-Earth or a volatile-rich sub-Neptune, especially near the 1.5--2.0~\rearth\ radius valley and the approximately 1.6~\rearth\ transition in rocky composition \citep{Rogers2015,Fulton2017}. For giant planets, mass, radius, age, and irradiation jointly constrain interior heavy-element content and thermal inflation, so dilution correction can substantially alter the inferred structure even when the broad giant-planet classification is unchanged \citep{Thorngren2016}.

\begin{figure}[t]
\centering
\includegraphics[width=\columnwidth]{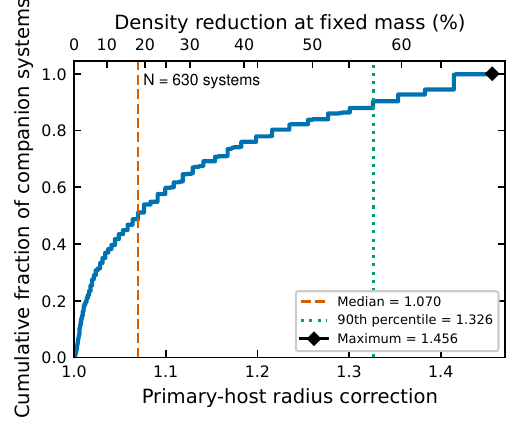}
\caption{Cumulative distribution of primary-host radius corrections for the 631 cumulative-survey hosts with at least one latest-epoch SOAR source inside 3 arcsec. The upper axis gives the corresponding density reduction if planet mass is unchanged. A 10\% radius increase reduces density by 25\%; the maximum 45.6\% radius correction reduces it by 67.6\%.}\label{fig:radiuscorr}
\end{figure}

Figure~\ref{fig:radiuscorr} shows the correction distribution. In the frozen table, the primary-host correction moves one planet upward across 1.6~\rearth, none across 2~\rearth, four across 4~\rearth, and 13 across 8~\rearth. These counts depend on the adopted empirical boundaries and current catalog radii.

If the planet transits a bound secondary, the correction is
\begin{equation}
\frac{R_{p,B}}{R_{p,0}}=
\frac{R_B}{R_A}\left(\frac{F_A+F_B}{F_B}\right)^{1/2}.
\label{eq:secondarycorr}
\end{equation}
One contrast measurement leaves both the secondary radius and physical association undetermined. We therefore release the measured contrasts and primary-host corrections, reserving secondary-host radii for systems with the necessary stellar characterization.

Equations~\ref{eq:primarycorr} and \ref{eq:secondarycorr} address third-light dilution while retaining the catalog stellar parameters used to convert transit depth into planet radius. In particular, Equation~\ref{eq:primarycorr} keeps the TIC radius of the presumed primary even when that radius came from blended photometry. Section~\ref{sec:deblendingfuture} describes the additional stellar and transit modeling required for a self-consistent correction.

Equation~\ref{eq:primarycorr} applies to planets already detected by TESS. A separate completeness effect arises when companion light dilutes a genuine transit below the detection threshold. For companion-to-primary flux ratio $f$, a primary-host transit depth is reduced by $(1+f)^{-1}$. Equal-brightness binaries halve the observed depth, whereas a $\Delta I=3$ companion reduces it by 5.9\%. This selection effect must be included when the observed companion deficit is converted to an intrinsic planet occurrence rate \citep{Bergsten2026}.

\section{Gaia Companion and Unresolved-binary Diagnostics}\label{sec:gaia}

A SOAR image reveals nearby sources, while Gaia adds the astrometric information needed to evaluate physical association and probes a wider separation range. We use Gaia common motion to extend the companion census and use astrometric and radial-velocity diagnostics to flag unresolved inner structure. Each observable retains its own selection function and interpretation.

\subsection{DR3 common-proper-motion search}

The common-motion search is intended to build a uniform wide-companion sample that can anchor the outer end of the separation distribution. Matching parallax and proper motion removes most chance alignments, while explicit angular, magnitude, and mass-ratio cuts let the same selection be applied to the simulated field population.

We performed a uniform Gaia DR3 search before constructing the combined companion census. We first identified each target's Gaia primary within 5 arcsec, ranking candidates by angular separation, parallax, and two-dimensional proper motion. All 2,730 coordinate-bearing targets received a preferred Gaia source, and none triggered the adopted ambiguity threshold. Of these, 2,600 had the distance and five-parameter astrometry required for the wider neighbor search.

For each primary, the Gaia distance estimate set the angular search radius corresponding to a projected separation of 10,000 au, $\rho_{\max}=10{,}000/d$ arcsec, subject to a 600 arcsec cap. We required the parallax difference to be below $3\sigma$, including a 0.15 mas systematic floor. Common motion was quantified as
\begin{equation}
\gamma_{\rm CPM}=\frac{|\boldsymbol{\mu}_A|+|\boldsymbol{\mu}_B|}
{|\boldsymbol{\mu}_A-\boldsymbol{\mu}_B|},
\end{equation}
with $\gamma_{\rm CPM}\geq5$. To retain close pairs with measurable orbital motion, we also required
\begin{equation}
\Delta\mu \leq 0.44\,{\rm mas\,yr^{-1}}
\left(\frac{\varpi}{\rm mas}\right)^{3/2}
\left(\frac{\rho}{\rm arcsec}\right)^{-1/2}
+2\sigma_{\Delta\mu},
\end{equation}
following the generous $5\,\msun$ envelope of \citet{ElBadry2021}. RUWE and duplicated-source flags are retained solely as diagnostics and play no role in rejecting pairs.

The final catalog contains 510 Gaia DR3 common-motion companions around 451 SOAR hosts after reciprocal searches of the same physical pair are collapsed. For the demographic comparison we treat this catalog as the wide-companion survey. A Robo-AO comparison found that Gaia DR2 recovered stellar pairs to about 1 arcsec at contrasts as large as six magnitudes \citep{Ziegler2018Gaia}; DR3 provides the longer-baseline astrometry used for the common-motion selection here \citep{Gaia2023}.

We define a deliberately simple working selection for the wide survey. Of the 1,199 demographic dwarf primaries, 1,179 have the Gaia five-parameter astrometry and photometry required for this comparison. We use Gaia from 300 to 5,000 au, require $\rho\geq1$ arcsec and $G_2\leq20.7$, and impose the same physical $q\geq0.4$ domain used for SOAR. For the Gaia detections, $q$ is estimated by inverting the Pecaut--Mamajek dwarf-sequence $M_G$--mass relation; the simulated field population is passed through the same angular, magnitude, and target-eligibility cuts. Within this 1,179-host Gaia-eligible subset, 88 common-motion companions satisfy the detection and mass-ratio criteria. Section~\ref{sec:shifted} therefore uses SOAR over 1--300 au and Gaia over 300--5,000 au in one separation-distribution likelihood. We use this DR3 selection for the wide-companion analysis. A source-level treatment of scanning law, crowding, and color-dependent Gaia completeness should be done with the final Gaia catalog.

The Gaia search serves two purposes: it provides likely physical associations for the released companion census and supplies the wide-separation comparison needed to distinguish a depleted close-binary population from a simple outward redistribution of the field distribution.

\subsection{Unresolved-binary diagnostics}\label{sec:gaiaflags}

Even a single Gaia catalog source can carry evidence of multiplicity. Comparing RUWE with \textsc{paired} shows how often astrometric and radial-velocity excess noise flag the same candidate inner systems and whether either reveals additional components in apparently wide hierarchies. We use the result as a diagnostic of unresolved structure rather than as another companion-frequency census.

We tested whether \textsc{paired}, the framework of \citet{Chance2025}, helps identify SOAR-resolved systems that Gaia leaves spatially unresolved. \textsc{paired} compares a source's reported Gaia RVS scatter with the noise distribution of stars at similar color and magnitude. Its $p_{\rm RV}$ is a single-star tail probability: the chance of obtaining at least that much radial-velocity noise under the adopted noise model. We use the authors' conservative $p_{\rm RV}<0.001$ excess-noise threshold and the exact source-level catalog join \citep{ChancePAIRED}. Injection tests show useful sensitivity primarily for semimajor axes below about 10 au, whereas RUWE$>1.4$ indicates a poor single-source astrometric fit over Gaia's observing baseline. Activity, variability, rotation, and calibration errors can affect either statistic \citep{Chance2025}.

The frozen ancillary diagnostic contains 630 cumulative-survey hosts with a latest-epoch SOAR component inside 3 arcsec. TOI-6094 is a false-positive target outside the demographic analysis and is excluded from this secondary test. We crossmatched the Gaia DR3 source associated with each of those primaries to the 2025 \textsc{paired} catalog. A Gaia primary match exists for 577 hosts, RUWE is available for 500, and \textsc{paired} covers 74. RUWE exceeds 1.4 for 282/500 (56\%), while 30/74 (41\%) of the RVS-covered hosts have $p_{\rm RV}<0.001$. Their different denominators preclude a direct performance comparison.

\begin{figure*}[t]
\centering
\includegraphics[width=0.98\textwidth]{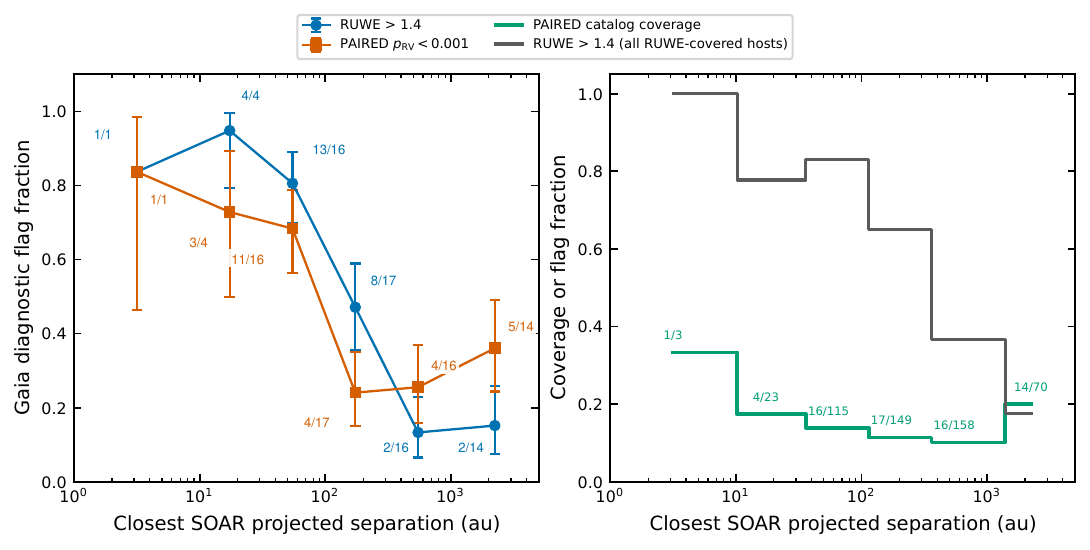}
\caption{Gaia diagnostics for the cumulative-survey hosts with a closest latest-epoch SOAR component inside 3 arcsec. Left: RUWE and \textsc{paired} flag fractions in projected-separation bins, restricted to the same \textsc{paired}-covered sources; every label gives the flagged count and the applicable denominator, and points show Jeffreys 68\% intervals. Right: \textsc{paired} catalog coverage and the RUWE$>1.4$ fraction among all RUWE-covered hosts. The bins include the 518 hosts with a distance-based projected separation, of which 68 have both diagnostics; the remaining 112 hosts lack a usable distance. At wide SOAR separations, an RV-noise flag most plausibly marks an additional unresolved subsystem rather than the imaged component.}\label{fig:paired}
\end{figure*}

Figure~\ref{fig:paired} compares the two diagnostics for the matched hosts. The matched comparison uses the 73 hosts with both statistics. The plotted separation bins contain 518 of the 630 hosts, including 68 of the 73 jointly covered hosts; the other 112 hosts, including five jointly covered hosts, lack the distance needed to convert the angular separation to au. RUWE flags 35/73 and \textsc{paired} flags 30/73; 20 have both flags, 15 are RUWE-only, 10 are \textsc{paired}-only, and 28 have neither. An exact McNemar test gives $p=0.424$. This paired-outcome test ignores the 20 both-flagged and 28 neither-flagged systems and asks whether the two discordant counts, 15 RUWE-only and 10 \textsc{paired}-only, differ more than expected if the diagnostics have the same marginal flag rate. ``Exact'' means that the probability is calculated from the finite binomial distribution rather than a large-sample approximation. The data are consistent with equal marginal flag rates. Among the 21 jointly covered hosts with a SOAR component inside 100 au, RUWE flags 18/21, \textsc{paired} flags 15/21, and their union flags 19/21. Thus \textsc{paired} adds one close-system flag, but its limited RVS coverage makes RUWE the more useful archive-wide screen for the resolved HRCam sample.

The complementarity is clearest for hierarchical candidates. Among 30 jointly covered hosts whose closest SOAR companion lies at 300--5,000 au, only four have RUWE$>1.4$, but nine satisfy the \textsc{paired} threshold and seven of those nine are not RUWE flagged. Because a companion hundreds of au away cannot generate the short-timescale RVS scatter to which \textsc{paired} is most sensitive, those seven are candidates for an additional unresolved inner companion rather than ``recoveries'' of the wide SOAR component.

\section{Close-binary Occurrence Analysis}\label{sec:suppression}

We now compare the TOI sample with the companions HRCam would detect around comparable field stars. Counts inside fixed projected separations quantify the close-companion deficit. A continuous model then tests how $S$ rises toward the field value, and the combined SOAR--Gaia distribution tests whether the missing close binaries have instead been redistributed to wider orbits. Subsequent planet-occurrence, stellar-mass, planet-radius, and disposition analyses explore the implications and limits of that primary companion-frequency result.

\subsection{Field model and completeness}

Target distance, binary contrast, and observing quality all shape the raw companion counts. Our benchmark therefore draws a field-binary population for every target and passes it through the same photometric, angular, mass-ratio, and survey-selection limits as the observations. The resulting forward model predicts the detectable companions for a TOI sample that follows the adopted field distribution.

\subsubsection{Baseline field population}

We compare the 1,199 dwarf primaries with the solar-type multiplicity census of \citet{Raghavan2010}. The primary sample is 1.83 times the 655-star culled solar-type sample used in the Paper II demographic analysis \citep{Ziegler2021}. Fractions of 0.33, 0.08, and 0.03 of systems were assigned one, two, and three companions. Periods follow a log-normal distribution with $\langle\log P({\rm day})\rangle=5.03$ and $\sigma_{\log P}=2.28$. The Monte Carlo proposal distribution is approximately uniform in mass ratio with a 10\% near-equal-mass excess; Section~\ref{sec:wideq} describes the event-level weights that convert it to the adopted mass- and separation-dependent field law. Binaries below the approximately 12-day tidal-circularization period were assigned $e=0$. Longer-period systems follow a Gaussian with mean 0.39 and standard deviation 0.31, truncated to $0\leq e\leq0.95$, as used for field binaries by \citet{GellerMathieu2012} and in the recent Hyades analysis of \citet{Torres2026}. The stored uniform draws are converted to this law by the exact density ratio, with random orbital phase and orientation retained.

Each simulated period was converted to a physical semimajor axis $a$ with Kepler's law. For a random mean anomaly $M$ and eccentricity $e$, we solved Kepler's equation for the eccentric anomaly $E$ and calculated the instantaneous three-dimensional separation $r=a(1-e\cos E)$. An isotropic viewing angle, represented by $\mu$ uniform on $[-1,1]$, then gives the sky-plane separation $s_{\rm proj}=r(1-\mu^2)^{1/2}$ and angular separation $\theta=s_{\rm proj}/d$. We applied the selected six-point contrast curve at $\theta$. Thus all comparisons with observed companions use projected separation, while the suppression laws below act on semimajor axis before projection; no constant conversion between $a$ and $s_{\rm proj}$ is assumed.

Projection statistically broadens a physical cutoff. In the target-sensitivity-convolved field simulation, the weighted median $s_{\rm proj}/a$ is 0.886, with a 16th--84th percentile range of 0.513--1.258. The ratio can exceed unity because eccentric systems spend more time near apastron. Among simulated detections with $s_{\rm proj}<58$ au, 80.9\% have $a<58$ au; conversely, 94.0\% of the detectable $a<58$ au population projects inside 58 au. For illustration, multiplying the intrinsic field-binary frequency by 0.15 for $a<58$ au produces an expected count at $s_{\rm proj}<58$ au equal to 0.313 of the unsuppressed expectation. Some wider systems project inward and some closer systems project outward, so a projected-separation ratio does not necessarily equal the imposed semimajor-axis multiplier.

\subsubsection{Matched photometry and HRCam detectability}

For each simulated target, the drawn component mass sets both stellar brightnesses. We interpolate the Pecaut--Mamajek dwarf sequence to obtain $M_{I,1}$ from $M_1$ and $M_{I,2}$ from $M_2=qM_1$, calculate $\Delta I=M_{I,2}-M_{I,1}$, and compare that contrast with the target's measured HRCam limit at the projected angular separation. The same flux ratio gives the combined-light magnitude-limited volume weight, $(1+10^{-0.4\Delta I})^{3/2}$ \citep{MoeKratter2021}. The observed conversion is the inverse operation: measured $\Delta I$ gives $M_{I,2}$ and hence empirical $q$. Both populations require $q\geq0.4$. This matched selection defines the main analysis; Section~\ref{sec:companions} reports a fixed $\Delta I\leq5.1$ cut only as a robustness check.

\subsubsection{Mass-ratio and catalog-selection calibration}\label{sec:wideq}

The Monte Carlo proposal treats all binaries with one simple mass-ratio law and initially gives every pair the combined-light advantage of an unresolved magnitude-limited system. Neither approximation is adequate across the SOAR--Gaia separation range. We therefore apply two event-level weights before filtering the simulation through HRCam or Gaia: one maps the proposal $q$ distribution onto the measured field distribution, and the other removes the companion's brightness boost whenever the target catalog would resolve the pair. The binary-suppression models are fitted only after those external population and selection weights are fixed.

\paragraph{Correction 1: resolved and unresolved targets.}

For an unresolved binary, the combined light allows the system to enter a magnitude-limited target list from a larger distance, so the combined-light volume factor in the preceding subsection is appropriate. Once Gaia catalogs the stars separately, however, the TOI is selected using the primary's catalog entry; the companion's light no longer extends the distance over which that primary enters the target sample. Giving a resolved pair the unresolved-system brightness boost would therefore overpredict the number of wide companions in the comparison sample.

Whether Gaia catalogs one source or two depends on the pair's angular separation and contrast, not directly on its projected separation in au. Empirical Gaia DR2 contrast-sensitivity maps show that faint companions must be farther from the primary to be recovered \citep{BrandekerCataldi2019}, and empirical wide-binary selection functions describe this close-pair recovery in angular-separation--contrast space \citep{ElBadryRix2019}. Gaia EDR3 validation likewise shows that close-pair completeness depends on angular separation, magnitude, and catalog processing \citep{Fabricius2021}. This treatment is especially relevant because the TIC target catalog was constructed from Gaia DR2 information \citep{Stassun2019}; Gaia--speckle comparisons also directly document the loss of close, high-contrast pairs \citep{Ziegler2018Gaia}.

We approximate the published Gaia DR2 50\%-recoverability contour by
\begin{equation}
\theta_{50}(\Delta G)=0.5+\left(\frac{\Delta G}{5}\right)^{2.5}\ \mathrm{arcsec},
\label{eq:gaiaangular}
\end{equation}
the inverse of $\Delta G=5(\theta-0.5)^{0.4}$. For consistency with the existing Gaia proxy in the forward simulation, we use the simulated $\Delta I$ as the contrast proxy. Each synthetic pair above this angular/contrast locus is treated as catalog-resolved before any binning in projected separation. Its original magnitude-limited weight contains its own combined-light factor
\begin{equation}
W_{{\rm blend},i}=\left(1+10^{-0.4\Delta I_i}\right)^{3/2}.
\label{eq:widesel}
\end{equation}
For a catalog-resolved pair we divide its weight by $W_{{\rm blend},i}$, returning that synthetic target to the primary-only selection weight; an unresolved pair retains its original weight. This pair-by-pair operation is preferable to one population-average factor because the brightness advantage ranges from large for nearly equal stars to negligible for a faint secondary. It removes only the extra target-selection volume assigned to that particular unresolved pair; Gaia detection completeness remains a separate selection effect. Because the sample spans a broad range of distances and contrasts, the angular boundary maps into a gradual change in projected-separation space, with different simulated pairs changing weight at different physical separations.

\paragraph{Correction 2: the field mass-ratio distribution.}

The field mass-ratio distribution depends on both primary mass and separation \citep{MoeDiStefano2017,ElBadryRix2019}. We use the median parameters in Table 2 of \citet{ElBadryRix2019}: five projected-separation intervals from 50 to 5,000 au, four primary-mass intervals from 0.4 to 2.5~\msun, a broken power law in $q$, and a separation-dependent twin excess above its fitted $q_{\rm twin}$. The exact 20 parameter sets and their implied fractions above $q=0.4$ are released electronically. Below 50 au, where that empirical table does not apply, we retain the approximately flat-plus-twin proposal law.

For every synthetic pair $i$ in the calibrated range, the mass-ratio importance weight is $w_{q,i}=p_{\rm EB}(q_i\mid M_{1,i},s_{{\rm proj},i})/p_{\rm sim}(q_i)$. This operation sets the relative weights of individual high- and low-$q$ binaries before the $q\geq0.4$ cut and before HRCam or Gaia selection. The complete population-and-selection weight is
\begin{equation}
f_i=f_{{\rm sel},i}\,w_{e,i}\,w_{q,i},
\label{eq:widefactor}
\end{equation}
where $w_{e,i}$ is the truncated-Gaussian eccentricity density divided by the uniform proposal density, and $f_{{\rm sel},i}=W_{{\rm blend},i}^{-1}$ for a catalog-resolved pair and unity otherwise. Only after these pair-level operations are the weights accumulated into projected-separation bins. Distance, primary mass, separation, contrast, and $q$ therefore enter each event explicitly. The matched Gaia range contains 69.6 expected companions under the resulting field model.

\paragraph{Field expectations across separation.}

Under the adopted eccentricity, mass-ratio, target-sensitivity, and catalog-selection model, the expected detectable companion counts are 84.8 inside 50 au, 124.4 inside 100 au, and 52.5 in the independent 100--300 au annulus. The corresponding observed counts are 24, 48, and 44. The El-Badry calibration applies from 50 au outward, while the angular resolved-target term changes gradually because the targets span a broad range of distance and contrast. These counts establish a strong close-companion deficit and a rise toward the field frequency in the 100--300 au annulus.

\subsubsection{Numerical implementation and convergence}

We used the full 10,000 Monte Carlo trials per target and an independent deterministic random stream for every TIC, so changing sample membership does not resimulate the remaining targets. Here ``Monte Carlo'' means repeated random draws from the field population, used to approximate the mean number and properties of companions that HRCam would detect. The multiplicity probabilities divide exactly into four 2,500-system blocks; each block contains 825, 200, and 75 one-, two-, and three-companion system draws per target. The stored uniform eccentricity and simple-$q$ draws serve only as proposal distributions; the density ratios in Equation~\ref{eq:widefactor} make their weighted distribution equal to the adopted field laws. An independent higher-resolution calculation with 100,000 trials per target changes the unweighted field expectations by only 0.02\% inside 50 au, 0.05\% inside 100 au, and 0.04\% in the 100--300 au annulus. Four independent blocks in the weighted calculation provide the convergence check for the adopted response.

As Section~\ref{sec:targetselection} shows, known multiples entered dedicated lists for repeat astrometry rather than as new survey systems. The cumulative archive is nevertheless heterogeneous in TOI class and repeat depth, so we retain fixed projected-scale limits of 50 and 100 au as the descriptive primary results. The independent 100--300 au annulus describes recovery. Section~\ref{sec:continuous} fits the shape only from 1 to 300 au and repeats the fit for targets first observed by the end of 2020.

\subsubsection{Likelihoods and model comparison}

For observed companion count $k$ and unsuppressed field expectation $E$, the model with relative companion frequency $S$ predicts the mean count $\lambda=SE$. We therefore use $k\sim{\rm Poisson}(\lambda)$. The likelihood $L(S)=P(k\mid S,E)$ is the probability of the observed count viewed as a function of $S$. We adopt a prior that assigns equal density to all nonnegative $S$; combining that prior with the likelihood gives the posterior, the probability distribution for $S$ after conditioning on these data and this model. In this case the posterior is $\mathrm{Gamma}(k+1,\mathrm{rate}=E)$. We quote its median and 16th--84th percentiles, which form a central 68\% credible interval. The maximum-likelihood estimate is $k/E$. We use $S_{50\,{\rm au}}\equiv S(s_{\rm proj}<50\,{\rm au})$ and $S_{100\,{\rm au}}\equiv S(s_{\rm proj}<100\,{\rm au})$ for the nested 50 and 100 au samples. A value $S=1$ means agreement with the specified field model, $S<1$ means fewer detected companions per host, and $S>1$ means an excess. The credible interval is conditional on that model; unmodeled selection effects remain outside it.

We also performed a target bootstrap: we drew 10,000 new samples of 1,199 hosts by sampling the observed target list with replacement. Each target's detections and expected count were carried together, so unusually sensitive or companion-rich hosts remained paired. The spread among these resamples measures sensitivity to which targets happen to enter the sample, rather than uncertainty in the adopted field model. Four disjoint simulation blocks provide an internal convergence test. Their fractional scatter in the full-sample expectation is 0.2\% inside 50 au and 0.3\% inside 100 au, below the counting and field-population uncertainties.

For comparisons among fitted separation models, we use the Akaike information criterion, ${\rm AIC}=2k_{\rm par}-2\ln\widehat{L}$, which balances goodness of fit against model complexity \citep{BurnhamAnderson2004}. Here $k_{\rm par}$ is the number of parameters fitted to these data, $L$ is the Poisson likelihood, and $\widehat{L}$ is its maximum. The ``candidate set'' is the finite list of step, smooth, constant, and shifted models compared in the relevant table. Lower AIC indicates less estimated information loss only within that list. As practical guidance, $\Delta{\rm AIC}<2$ denotes comparable support, values near 4--7 indicate appreciably less support, and values above 10 indicate little support. AIC supplies neither an absolute goodness-of-fit measure nor a model probability.

The smooth recovery law represents a gradual transition rather than a hard reduction below one physical separation. Its transition-width parameter allows suppression to recover over the range expected from disk truncation, disk dispersal, and later dynamical mechanisms. The AIC penalty requires that this added flexibility improve the projected-data likelihood enough to justify the additional parameter.

In summary, the forward model compares like with like, using the same empirical $q\geq0.4$ domain, orbital projection, and target-specific detectability, while its wide normalization accounts for the separation dependence of selection and the field mass-ratio distribution. The fixed-scale deficit and fitted recovery shape can therefore be interpreted relative to one calibrated field benchmark.

\subsection{Fixed-separation companion frequencies}

Fixed limits of 50 and 100 au provide the most transparent view of the close-source deficit before any recovery function is chosen. At each limit, the observed count is compared with the field population after projection through the sensitivity of the actual survey.

Physical field orbits are projected before comparison with the observed histogram. The fixed limits below are therefore defined in apparent projected separation, whereas a step amplitude refers to semimajor axis.

\subsubsection{The close-scale deficit}

Within the 1,199-host demographic sample and inside an apparent projected separation of 50 au, we observe 24 candidate visual companions and expect 84.8 detectable field companions. The posterior relative companion frequency is
\begin{equation}
S_{50\,{\rm au}}=0.291^{+0.062}_{-0.054}.
\end{equation}
Within the same demographic sample and inside 100 au, we observe 48 sources and expect 124.4 field companions, giving
\begin{equation}
S_{100\,{\rm au}}=0.391^{+0.058}_{-0.053}.
\end{equation}
The cumulative 1,199-host demographic sample therefore contains 3.5 and 2.6 times fewer detected nearby sources than the field-binary expectation at the two limits. The target bootstrap likewise retains a substantial deficit at both limits. Chance projections, if present, add to the observed counts and therefore weaken rather than generate this deficit, although association follow-up remains necessary for physical interpretation.

Omitting the 66 hosts without catalog masses would introduce a substantial bias in the companion statistics. The temperature-based masses assigned to these hosts retain 17 of the 24 observed sources inside 50 au and 40 of the 48 inside 100 au. Deleting them would leave only 7 and 8 sources while removing merely 3.4 and 6.3 expected field detections. Availability of mass estimates in the TIC is therefore strongly correlated with resolved multiplicity. We retain these hosts with explicit, flagged mass estimates rather than turn a missing catalog value into an outcome-dependent sample cut.

The fixed-scale suppression inferred with this model is comparable to the $0.34^{+0.14}_{-0.15}$ result from \citet{Kraus2016}, although the samples and separation definitions differ, and remains qualitatively consistent with the TESS speckle distributions of \citet{Howell2021} and \citet{Lester2021}. The direct occurrence calculation of \citet{Sullivan2026} finds 50--58\% fewer Kepler planets in its adopted binary samples, but it explicitly models transit completeness and therefore measures a different quantity.

\subsubsection{Recovery in the 100--300 au annulus}

The frequency rises rather than returning immediately to the field rate at 100 au. In the independent 100--300 au annulus of the same 1,199-host demographic sample, 44 sources are observed and the field model predicts 52.5, giving $S_{100-300}=0.851$ (0.731--0.984). This prediction applies the adopted eccentricity and El-Badry population weights, the target-specific contrast curves, and removal of each catalog-resolved pair's own combined-light weight. We retain the annulus as a descriptive result because the cumulative TOI and repeat-observation selection is heterogeneous.
Table~\ref{tab:suppression} gives the fixed 50 and 100 au results for the full sample and the principal subsamples.

\begin{deluxetable*}{lrrrrr}
\tablecaption{Close-companion frequency at fixed separation limits\label{tab:suppression}}
\tablehead{\colhead{Sample}&\colhead{$N_\star$}&
\multicolumn{2}{c}{$s_{\rm proj}<50$ au}&\multicolumn{2}{c}{$s_{\rm proj}<100$ au}\\
\colhead{}&\colhead{}&\colhead{$N_{\rm obs}/N_{\rm field}$}&\colhead{$S$ (16--84\%)}&
\colhead{$N_{\rm obs}/N_{\rm field}$}&\colhead{$S$ (16--84\%)}}
\startdata
Mass-matched hosts &1,199&24/84.8&$0.291\ (0.236$--$0.353)$&48/124.4&$0.391\ (0.338$--$0.449)$\\
Confirmed or known planets &291&1/27.7&$0.061\ (0.026$--$0.119)$&1/37.4&$0.045\ (0.019$--$0.088)$\\
$T<13$ &1188&24/84.7&$0.291\ (0.237$--$0.353)$&48/124.1&$0.392\ (0.339$--$0.451)$\\
$d<500$ pc &986&23/83.2&$0.284\ (0.230$--$0.346)$&47/117.7&$0.405\ (0.350$--$0.466)$\\
$4500<T_{\rm eff}<6500$ K &1005&22/69.1&$0.328\ (0.264$--$0.402)$&43/103.1&$0.424\ (0.363$--$0.491)$\\
\enddata
\tablecomments{Each sample appears once; the paired column groups show the nested 50 and 100 au tests. $S_{50\,{\rm au}}$ and $S_{100\,{\rm au}}$ are the posterior median resolved-source frequencies inside 50 and 100 au, respectively, relative to the sensitivity-convolved physical field-binary model. ``Mass-matched'' denotes the primary-mass range over which observed and simulated contrasts are converted consistently to $q$. A host enters the confirmed/known row if any retained signal has a CP or KP disposition. Physical association is not required for the observed SOAR count.}
\end{deluxetable*}

Taken together, the fixed-separation analysis establishes the primary demographic result without depending on a fitted transition shape: the TOI sample contains only about one third of the field expectation inside 100 au, with a clear rise toward larger separation.

\subsection{M-dwarf hosts}\label{sec:mdwarfs}

Lower-mass primaries require their own comparison population because the main field model is calibrated for solar-type stars. An exploratory M-dwarf calculation using a Winters-based benchmark tests whether the qualitative close-source deficit extends to this regime; its amplitude can only be interpreted relative to that separate field sample.

For M-dwarf hosts we use the nearby-star statistics of \citet{Winters2019} rather than extending the Raghavan solar-type model below its calibrated mass range. The M-dwarf sample applies the same retained-signal, $R_p\leq15$~\rearth, contrast-curve, known-distance, and brown-dwarf-vetting requirements as the primary sample, but selects $2400\leq T_{\rm eff}<3900$ K. It contains 89 hosts at a median distance of 43.0 pc.

The field model uses the Winters et al. multiplicity rate of $26.8\pm1.4\%$ and their cataloged angular separations. We converted the 283 companion rows with finite parallaxes to projected au; their median is 17.6 au. This empirical coordinate is already the observed sky-plane separation, including the catalog's treatment of unresolved astrometric and spectroscopic systems, so we resample it directly. Converting it to an assumed semimajor-axis distribution and projecting it again would add assumptions without adding information.

For each target we drew 20,000 field systems. Stellar mass ratios were drawn from the approximately uniform high-$q$ distribution reported by \citet{Winters2019}, with the minimum set by the hydrogen-burning limit for that primary. The Pecaut--Mamajek dwarf sequence converted $q$ to $\Delta I$; the target distance converted projected separation to angle; and the measured HRCam curve determined whether the companion would be detected. We applied the same combined-light volume weight as in the solar-type analysis. The separation distribution and contrast conversion are specific to M dwarfs.

\begin{figure}[t]
\noindent\includegraphics[width=\columnwidth]{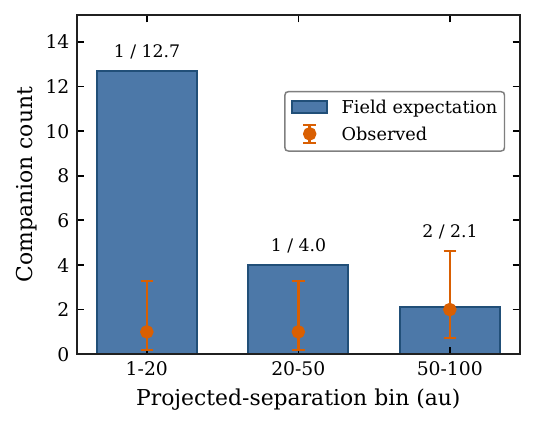}
\caption{Exploratory M-dwarf companion test using the Winters nearby-star field population. Bars show the sensitivity-convolved field expectation in three independent projected-separation bins, and orange points show the observed SOAR counts with 68\% Poisson intervals. Labels give observed/expected counts. The deficit is present throughout the plotted range, but the small sample and different field benchmark preclude a calibrated amplitude comparison with the solar-type result.}\label{fig:mdwarf}
\end{figure}

The exploratory result shows a deficit relative to this adopted M-dwarf field model (Table~\ref{tab:mdwarf}; Figure~\ref{fig:mdwarf}). One source is observed inside 20 au where 12.7 are expected, two inside 50 au where 16.7 are expected, and four inside 100 au where 18.8 are expected. The posterior relative frequencies are $0.132$ (0.056--0.258), $0.160$ (0.082--0.277), and $0.249$ (0.152--0.381). Propagating the published 1.4 percentage-point uncertainty in the multiplicity rate changes the intervals negligibly compared with Poisson counting uncertainty.

\begin{deluxetable*}{crrrrrr}
\tablecaption{M-dwarf close-companion suppression\label{tab:mdwarf}}
\tablehead{\colhead{$s_{\rm proj}$ limit}&\multicolumn{3}{c}{All stellar companions}&\multicolumn{3}{c}{$q\geq0.4$ matched}\\
\colhead{}&\colhead{$N_{\rm obs}$}&\colhead{$N_{\rm field}$}&\colhead{$S$ (16--84\%)}&\colhead{$N_{\rm obs}$}&\colhead{$N_{\rm field}$}&\colhead{$S$ (16--84\%)}}
\startdata
20 au &1&12.7&$0.132\ (0.056$--$0.258)$&1&11.7&$0.144\ (0.061$--$0.281)$\\
50 au &2&16.7&$0.160\ (0.082$--$0.277)$&1&15.1&$0.111\ (0.047$--$0.217)$\\
100 au&4&18.8&$0.249\ (0.152$--$0.381)$&2&16.9&$0.158\ (0.081$--$0.274)$\\
\enddata
\tablecomments{The all-stellar columns use the Winters nearby-star field population. The matched columns use the same $q\geq0.4$ domain as the solar-type calculation. Because the Winters catalog and Raghavan forward model are different comparison populations, even the equal-$q$ results lie on distinct field baselines and permit only a qualitative cross-mass comparison.}
\end{deluxetable*}

The result agrees qualitatively with the dearth of close sources among M-dwarf TOIs reported by \citet{Clark2022}, while the small sample and distinct field model leave the cross-mass suppression law weakly constrained. Restricting both calculations to $q\geq0.4$ gives conditional two-sided probabilities of $p=0.163$ inside 50 au and $p=0.167$ inside 100 au under one common relative-frequency factor. For this and the subgroup tests below, a conditional two-sided $p$-value is the probability, assuming the two groups share one underlying relative frequency and holding the total number of detections fixed, of obtaining a split at least this unequal in either direction. The small M-dwarf counts and the different field populations dominate this comparison.

The M-dwarf subset extends the qualitative close-source deficit beyond the 1,199 solar-type dwarfs. Measuring how its amplitude changes with stellar mass will require a larger confirmed-planet sample and a completeness-calibrated M-dwarf control survey.

\subsection{Continuous recovery with separation}\label{sec:continuous}

The fixed-separation counts reveal a deficit and a rise toward larger separations, but they leave the transition scale and width unspecified. We model that recovery in physical semimajor axis and project it through the survey response, preserving the distinction between orbital scale and observed sky-plane separation.

The fixed annuli establish a deficit and its recovery without fitting a cutoff. We model the gradual rise with a phenomenological logistic transition while retaining the same field population, orbital projection, magnitude-limited weighting, and contrast selection. This form is monotonic and bounded, with an inner level, midpoint, and width. We define it in $\log_{10}a$ because binary and disk scales span orders of magnitude, so a width in dex describes a fractional separation range. Each simulated binary's field weight is multiplied by
\begin{equation}
S(a)=S_0+\frac{1-S_0}
{1+\exp\{-[\log_{10}(a)-\log_{10}(a_{50})]/w\}},
\label{eq:recovery}
\end{equation}
Here $S_0$ is the inner asymptote, $a_{50}$ is the midpoint where $S=(1+S_0)/2$, and $w$ is the transition width in dex. The outer asymptote is fixed at unity because $S(a)$ is defined relative to the field and HRCam alone cannot constrain both a wide-separation normalization and a recovery scale. Therefore, $a_{50}$ is a model-dependent midpoint toward an assumed field-normalized outer population, not a boundary read directly from the images.

\begin{figure*}
\centering
\includegraphics[width=\textwidth]{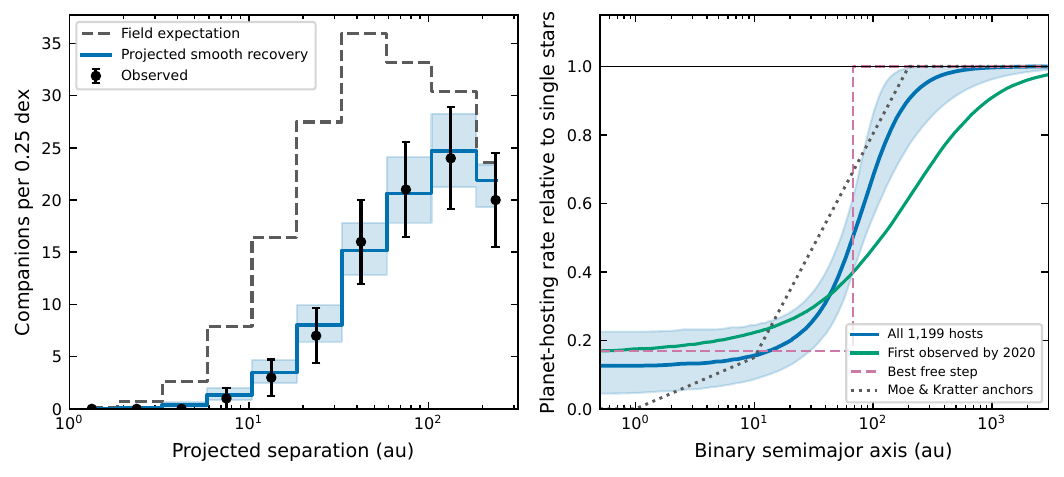}
\caption{Forward-projected separation-recovery fit. Left: observed projected-separation counts, the field expectation, and posterior-predictive recovery counts from Equation~\ref{eq:recovery}. For every posterior draw, the law weights binaries in true semimajor axis before orbital projection and target-specific HRCam selection; the curve and band summarize the resulting expected projected counts. Right: the conditional recovery multiplier as a function of binary semimajor axis. The band is the 16th--84th percentile for all 1,199 demographic hosts, giving $a_{50}=74$ au (57--101 au). The green curve is the weakly constrained pre-2021 check, the magenta line is the best-fit step, and the gray dotted curve is the literature benchmark from Paper II. The step and smooth fits have comparable AIC ($\Delta{\rm AIC}=1.7$).}\label{fig:continuous}
\end{figure*}

The trial populations were projected through the target-specific HRCam response and fit to the observed 0.25 dex projected-separation bins with a Poisson likelihood. We evaluated the three-parameter posterior on a discrete grid and used its marginalized distributions for the reported intervals and predictive band. The ``posterior-predictive recovery counts'' in the left panel of Figure~\ref{fig:continuous} are constructed in physical semimajor axis: each posterior draw weights binaries through Equation~\ref{eq:recovery}, after which orbital projection and the individual HRCam contrast curves produce expected counts in projected-separation bins. The blue curve and band summarize those expected counts over the posterior. Appendix~\ref{app:fitdetails} gives the response-matrix equation, likelihood, priors, parameter grids, and posterior-sampling procedure.

After applying the event-level eccentricity, mass-ratio, and resolved-target weights, the posterior gives $S_0=0.120$ (0.038--0.222), $a_{50}=74$ au (57--101 au), and $w=0.194$ dex (0.077--0.337 dex). These are parameters of the semimajor-axis forward model, not direct projected-separation measurements. A prior uniform rather than log-uniform in $w$ gives $S_0=0.095$, $a_{50}=76$ au, and $w=0.275$ dex; the midpoint is stable to this prior change. The response changes gradually with projected separation because population and detection weights are applied to individual simulated pairs.

We compared the smooth law with a fitted step and a constant factor over 1--300 au using the same response and Poisson likelihood. The published Paper II step is plotted separately as a literature benchmark and is not included among the models fitted here. The best step-cutoff model has $S_{\rm bin}=0.169$ inside $a_{\rm cut}=68$ au, close to the Paper II result of $S_{\rm bin}=0.15$ inside 58 au \citep{Ziegler2021}. This agreement persists despite the expanded sample and target-level response model. The fitted step is worse than the smooth law by only $\Delta{\rm AIC}=1.7$; the constant model is worse by 20.8. Thus the data require recovery with separation relative to a constant factor, while the present sample does not decisively distinguish a smooth transition from a sharp step. The comparison grids are specified in Appendix~\ref{app:fitdetails}.

The pre-2021 robustness subset contains 440 of the 1,199 demographic hosts and only 28 matched sources from 1--300 au, too few to constrain the shape. The smooth-law midpoint is $164$ au (81--418 au), much of whose interval lies beyond the fitted range; the constant and step descriptions are within $\Delta{\rm AIC}=2.8$. This cohort confirms that the deficit predates multiplicity-based repeat visits, while its recovery scale remains weakly determined.

The scale is qualitatively consistent with the weak effect beyond about 200 au inferred by \citet{MoeKratter2021} and the reduced pebble-fed planet formation below about 160 au predicted by \citet{Venturini2026}. These studies provide contextual benchmarks rather than sample-specific predictions. The HRCam data favor gradual recovery over one hard cutoff, but the midpoint remains model dependent.

\subsection{Suppressed versus shifted binary distributions}\label{sec:shifted}

Two population histories can produce a shortage of close companions: fewer binaries among detected TOIs, or a redistribution that moves the same binaries outward. Combining SOAR with Gaia supplies the separation leverage to distinguish them by checking whether the missing close systems reappear at wider separations.

\subsubsection{Competing population models}

Earlier TESS speckle surveys found a characteristic binary scale near 100 au rather than the 40--50 au field scale \citep{Howell2021,Lester2021}. This suggests a distinct alternative to suppression: the field distribution could move outward while preserving total multiplicity. The Gaia survey provides the wide-separation leverage needed to test that redistribution.

The comparison uses the same $q\geq0.4$ dwarf-primary population: 92 SOAR sources from 1--300 au and 88 Gaia common-motion companions from 300--5,000 au among 1,179 searchable primaries. Simulations apply the target-specific HRCam curves, the Gaia cuts $\rho\geq1$ arcsec and $G_2\leq20.7$, the angular resolved-target weighting in Equation~\ref{eq:gaiaangular}, and the event-level eccentricity and El-Badry $q$ weights. Suppression models weight the Raghavan population with a step or smooth law. Shifted models change the log-period center and, where allowed, its width while conserving multiplicity. A hybrid also allows the 100 au distribution's amplitude to vary. The same externally determined response is used for every model in the comparison.

We fit all models to the same 0.25 dex bins over 1--5,000 au with a Poisson likelihood after applying the matching SOAR and Gaia selection matrices. This comparison uses maximum-likelihood parameters for AIC, rather than the HRCam-only posterior medians in Section~\ref{sec:continuous}. Appendix~\ref{app:fitdetails} gives the optimizer, parameter bounds, repeatability settings, and AIC construction.

\subsubsection{Combined SOAR--Gaia results}

\begin{figure*}[t]
\centering
\includegraphics[width=0.96\textwidth]{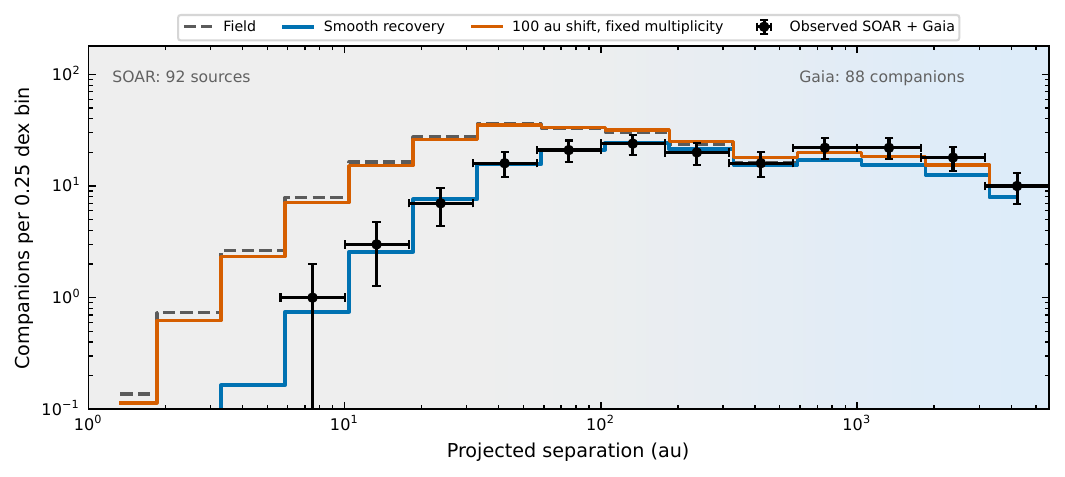}
\caption{Combined SOAR--Gaia comparison of suppression and the literature-motivated 100 au shifted distribution. The points are exactly the 92 SOAR sources from 1--300 au and 88 Gaia companions from 300--5,000 au used in the likelihood, all in the empirical $q\geq0.4$ domain. The dashed curve is the field expectation, the blue curve is the best smooth-recovery model, and the orange curve preserves field multiplicity while shifting the characteristic binary scale to 100 au. The background color shows the smooth survey handoff rather than a hard boundary. At each separation-bin range it is set by the fraction of companions contributed by SOAR and Gaia, with interpolation used to smooth the transitions between bins; gray and blue are the two endpoints.}\label{fig:shifted}
\end{figure*}

The smooth model has the lowest AIC (Table~\ref{tab:shiftmodels}; Figure~\ref{fig:shifted}). The audited Gaia bins contain 88 companions where the pairwise-corrected field model predicts 69.6, a descriptive observed-to-field ratio of 1.26. For a fixed Poisson mean this excess has a one-sided probability of 0.019; a 10\% upward field-normalization shift raises the expectation to 76.6 and the tail probability to 0.108. After accounting for uncertainties in the field normalization and Gaia selection, the wide population is consistent with the model, but we cannot claim exact agreement. The close SOAR bins remain strongly deficient. The combined maximum-likelihood smooth law gives $S_0=0.030$, $a_{50}=64$ au, and $w=0.256$ dex; these values differ from the HRCam-only posterior because this fit includes the Gaia leverage and reports a maximum rather than marginalized intervals.

The best step has $S=0.169$ below $a=70$ au and is $\Delta{\rm AIC}=1.3$ above the smooth law. Thus the combined data establish an inner deficit followed by wide recovery but do not distinguish sharply between smooth and step-function recovery. A freely shifted, field-width, fixed-multiplicity model drives its mean to the upper search bound, $\log P=7.0$, yet remains worse than the smooth model by $\Delta{\rm AIC}=16.5$; allowing its width to vary does not improve the likelihood. The fully specified 100 au shift is worse by 60.2. Allowing that distribution to reduce its amplitude gives $A=0.683$, retains the field width, and remains worse by $\Delta{\rm AIC}=35.9$. It is also no longer a pure redistribution model because its amplitude removes companions.

The combined SOAR--Gaia result therefore favors a close companion deficit followed by recovery toward a field-compatible wide population, rather than a multiplicity-conserving shift of the entire field binary distribution. The normalization and its remaining Gaia/TOI limitations are addressed in Section~\ref{sec:discussion}.

\begin{deluxetable*}{lrrrl}
\tablecaption{Combined SOAR--Gaia binary-distribution models over 1--5,000 au\label{tab:shiftmodels}}
\tablehead{\colhead{Model}&\colhead{$k_{\rm par}$}&\colhead{AIC}&\colhead{$\Delta$AIC}&\colhead{Key fitted result}}
\startdata
Smooth suppression recovery &3&59.9&0.0&$S_0=0.030$, $a_{50}=64$ au, $w=0.256$ dex\\
Suppressed step &2&61.2&1.3&$S=0.169$ for $a<70$ au\\
Freely shifted, field width, fixed multiplicity &1&76.4&16.5&mean reaches $\log P=7.0$ bound\\
Freely shifted and narrowed, fixed multiplicity &2&78.4&18.5&mean and width return to field-width bound\\
100 au shift, free amplitude and width &2&95.8&35.9&$A=0.683$; width returns to field value\\
100 au shift, field width and multiplicity &0&120.0&60.2&fully specified redistribution\\
Pairwise-corrected Raghavan field &0&129.2&69.3&fixed field benchmark\\
\enddata
\tablecomments{All likelihoods use the same mass-ratio-matched dwarf-primary population, SOAR from 1--300 au, Gaia DR3 common-motion companions from 300--5,000 au, and the corresponding survey response defined in Section~\ref{sec:suppression}. $k_{\rm par}$ counts parameters fitted to these data. The 100 au center is fixed using a representative $1\,\msun$ primary and $q=0.6$ companion. The free-amplitude row is a suppression--redistribution hybrid rather than a multiplicity-conserving alternative.}
\end{deluxetable*}

\subsection{Implications for planet occurrence rates}\label{sec:occurrence}

A wide-field transit survey selects systems with at least one detectable transit. This makes selection different for single and binary stars. A single star provides one possible planet host, whereas a binary provides two. If planets were equally common and equally detectable around both components, and detections were rare, a binary could be nearly twice as likely to enter the TOI sample. Binaries would then be overrepresented even if they had no effect on planet formation.

The actual enhancement is smaller than a factor of two. A faint secondary contributes less light, making its planets harder to detect in the blended photometry. Light from either star also dilutes transits around the other, lowering their signal-to-noise ratio and sometimes hiding them altogether. These effects bias planet radii, detection completeness, and occurrence rates when unresolved companions are not modeled \citep{Ciardi2015,Bouma2018,MoeKratter2021,Bergsten2026}. Catalog construction, centroid tests, and follow-up vetting also affect which star can supply a retained candidate. Thus, the second potential host raises the chance of detecting a planet, while dilution lowers it. The balance depends on mass ratio, angular separation, planet properties, and the TESS aperture.

Our companion counts and fitted $S_0$, $a_{50}$, and $w$ describe the selected TOI sample regardless of this interpretation. Because TOI selection may favor binaries, the observed close-binary deficit likely understates the physical suppression. Below, we describe this two-star selection effect and then estimate its population impact assuming that every retained planet orbits the catalog primary.

\subsubsection{Multiple host bias}

Previous occurrence studies treat unresolved binaries by assigning planets and detection completeness separately to the primary and secondary stars \citep{Bouma2018,Sullivan2026}. We use a simple model to show how this affects our companion-frequency measurement. Let $D$ mean that a system enters the retained TOI sample. For a binary with catalog primary A and secondary B, let $p_A$ and $p_B$ be the probabilities that each star would produce a retained signal if binarity did not change planet occurrence. These probabilities include transit geometry, dilution, TESS noise and aperture, pipeline recovery, centroid tests, and follow-up disposition. Let $r(a,q)$ be the factor by which a binary with semimajor axis $a$ and mass ratio $q$ changes planet occurrence around each star. If the two stars provide independent chances to enter the sample, the probability that at least one does so is
\begin{equation}
P(D\mid A,B)=1-[1-r(a,q)p_A][1-r(a,q)p_B].
\label{eq:twohostselection}
\end{equation}

The first bracket in Equation~\ref{eq:twohostselection} is the probability that primary A does not produce a retained signal, and the second is the same probability for secondary B. Multiplying them gives the probability that neither star produces a signal. Subtracting that product from one gives the probability that at least one star does. The factor $r(a,q)$ appears in both brackets because binarity can change planet occurrence around both stars. The terms $p_A$ and $p_B$ describe how likely we are to detect a planet around each one. A second star raises the chance of entering the sample, but its faintness and transit dilution can lower both detection probabilities. Therefore, no single correction applies to every binary, and the enhancement can be less than two even when both stars lie within the aperture.

Here, ``rare'' means that $r(a,q)p_A\ll1$ and $r(a,q)p_B\ll1$, so the product term $r(a,q)^2p_Ap_B$ can be ignored. This is typical for wide-field transit surveys, in which only a small fraction of searched stars yield detected planet candidates. Equation~\ref{eq:twohostselection} then reduces to $r(a,q)(p_A+p_B)$. Relative to a primary-only calculation, the extra chance supplied by the secondary can be written as
\begin{equation}
\mathcal{E}_{\rm host}=1+\frac{p_B}{p_A},
\label{eq:opportunityboost}
\end{equation}
where $\mathcal{E}_{\rm host}\simeq1$ if the secondary cannot supply a detectable planet and $\mathcal{E}_{\rm host}\simeq2$ if planets are equally detectable around both stars. Most systems fall between these limits, depending on their contrast, stellar properties, dilution, planet population, and observing history.

TESS pixels are approximately 21 arcsec wide, so both stars can contribute signals in many pairs separated by at least 10 arcsec. Difference-image centroids, ground-based time-series photometry, and other follow-up can locate the transit or reject an eclipsing-binary blend. However, they do not remove this selection effect because a planet transiting the secondary is still real. A complete correction must model both stars rather than apply a universal factor of two.

Here $S$ denotes the companion-frequency ratio measured in the selected TOI sample, whereas $r$ is the planet-occurrence factor for each star in a binary relative to a comparable reference star whose planet occurrence is not reduced by a close companion. The reference can be a single star or a component of a sufficiently wide binary for which the suppression has recovered to the field level. The factor $\mathcal{E}_{\rm host}$ accounts for the additional opportunity that either component can supply the detected planet. Before accounting for how binaries enter the selected sample, these quantities satisfy $S\simeq r\,\mathcal{E}_{\rm host}$, or
\begin{equation}
r\simeq\frac{S}{\mathcal{E}_{\rm host}}.
\label{eq:rdeboost}
\end{equation}
Thus, $r$ is the companion-frequency ratio corrected for the extra chance that either star can provide the detected planet. For example, inside 50 au we measure $S=0.291$. If only the primary can provide a detectable planet, $\mathcal{E}_{\rm host}=1$ and $r=0.291$, corresponding to approximately 71\% lower planet occurrence per star than for the reference population. If both stars provide equal detection opportunities, $\mathcal{E}_{\rm host}=2$ and $r=0.146$, corresponding to approximately 85\% lower occurrence per star. Inside 100 au, the corresponding factors are $r=0.391$ for $\mathcal{E}_{\rm host}=1$ and $r=0.196$ for $\mathcal{E}_{\rm host}=2$, or approximately 61\% and 80\% lower occurrence per star, respectively. These are limiting-case interpretations rather than occurrence measurements. They show only how the assumed contribution from the secondary changes the result. A full correction requires the separate detection probabilities in Equation~\ref{eq:twohostselection}.

\subsubsection{Primary-only population illustration}

A direct occurrence calculation requires a parent sample of all searched stars, transit completeness for binaries, candidate reliability, and the probability that the planet orbits either star. For comparison with earlier work, we also give a simpler primary-only estimate corresponding to the $\mathcal{E}_{\rm host}=1$ scenario above, in which only the catalog primary supplies a detectable planet. It counts one possible planet host per field system and treats Equation~\ref{eq:recovery} as that system's planet yield relative to a comparable single star or wide binary. If the true $\mathcal{E}_{\rm host}$ lies between one and two, then $r=S/\mathcal{E}_{\rm host}<S$, so the per-star suppression and the resulting reduction in population-wide planet occurrence would be stronger than the primary-only values presented below. Calculating the size of that additional reduction requires a completeness model that includes both stars. A volume-limited calculation counts systems within a fixed volume without weighting them by brightness. A magnitude-limited calculation gives more weight to bright unresolved multiples because they can enter a flux-limited catalog from farther away.

We simulated solar-type field systems with the same ingredients used in the forward model: single, binary, triple, and quadruple probabilities of 0.56, 0.33, 0.08, and 0.03; the adopted period and mass-ratio distributions; and the 10\% twin component \citep{Raghavan2010}. In higher-order systems, the closest companion in the selected mass-ratio range sets $S(a)$. This assumes that the closest relevant companion dominates rather than multiplying several uncertain suppression effects. We integrate Equation~\ref{eq:recovery} directly over semimajor axis. Projected separation enters only when fitting the model to the observations. If $\eta_0$ is the planet occurrence for an unaffected system, the population average is
\begin{equation}
\frac{\eta_{\rm all}}{\eta_0}
= \sum_m p_m \left\langle S(a_{\min})\right\rangle_m
=1-F_{\rm lost},
\label{eq:occurrenceimpact}
\end{equation}
Here, $m$ is the multiplicity class, $p_m$ is its field frequency, and the angle brackets give the average $S(a_{\min})$ for simulated systems in that class. A system without a relevant companion has $S=1$. For a visible-light magnitude-limited sample, the empirical component luminosities give the combined-light volume weight $(1+\sum_j10^{-0.4\Delta I_j})^{3/2}$ \citep{MoeKratter2021}. We calculate Equation~\ref{eq:occurrenceimpact} for every draw from the recovery-law posterior.

\subsubsection{Primary-only yield implications and limitations}

Under the primary-only assumption, the fitted model reduces planet yield by $F_{\rm lost}=0.170$ (0.155--0.183) in a volume-limited population within the measured $q\geq0.4$ range and by 0.244 (0.222--0.262) in a magnitude-limited sample. Extending the same separation law to companions with $0.1\leq q\leq1$ gives reductions of 0.233 (0.212--0.251) and 0.301 (0.274--0.323). The calculation keeps the Raghavan multiplicity fractions because it describes the underlying field population, not the resolved-target sample. Table~\ref{tab:occurrenceimpact} separates the measured and extrapolated cases.

\begin{deluxetable*}{llrrrr}
\tablecaption{Primary-only estimates of the planet-yield reduction\label{tab:occurrenceimpact}}
\tablehead{
\colhead{Mass-ratio domain}&\colhead{Selection}&
\colhead{$f_{\rm mult}$}&\colhead{$F_{\rm lost}$ (16--84\%)}&
\colhead{$\eta_{\rm all}/\eta_0$}&\colhead{$\eta_0/\eta_{\rm all}$}
}
\startdata
Measured $q\geq0.4$ & Volume limited &0.333&$0.170\ (0.155$--$0.183)$&0.830&1.20\\
Measured $q\geq0.4$ & Magnitude limited &0.465&$0.244\ (0.222$--$0.262)$&0.756&1.32\\
Extrapolated $0.1\leq q\leq1$ & Volume limited &0.440&$0.233\ (0.212$--$0.251)$&0.767&1.30\\
Extrapolated $0.1\leq q\leq1$ & Magnitude limited &0.552&$0.301\ (0.274$--$0.323)$&0.699&1.43\\
\enddata
\tablecomments{$f_{\rm mult}$ is the fraction of field systems with at least one companion in the stated mass-ratio range after selection weighting. The intervals include uncertainty in the recovery law but hold the field multiplicity distribution fixed. The all-$q$ rows assume that the same separation law applies below the survey's $q$ limit. Each row counts only one possible planet host per system, so these values are not per-star occurrence rates or direct TESS occurrence measurements. Modeling both stars requires the separate selection terms in Equation~\ref{eq:twohostselection}.}
\end{deluxetable*}

The direct Kepler calculation of \citet{Sullivan2026} provides a closer measure of $\eta_{\rm binary}/\eta_{\rm single}$. After modeling transit completeness and testing both primary and secondary hosts, they find 50--58\% fewer planets per system in their selected close-binary samples.

Our primary-only estimate shows that the companion deficit is large enough to affect overall planet yields, but the values are not per-star occurrence rates. Measuring those rates requires a full transit-completeness calculation with planets injected around both stars.

\subsection{Planet-radius dependence}\label{sec:radiussplit}

Disk truncation and transit dilution may leave different signatures across planet size. Dividing systems by their largest retained catalog planet or candidate radius provides an exploratory test of that dependence, with candidate reliability, distance, and dilution treated as correlated alternatives.

\subsubsection{Sample definition and close-scale response}

A resolved source is a host-level property, so each TIC enters the planet-radius comparison once using its largest retained planet or candidate. Dividing the sample at 9~\rearth\ gives 638 smaller-planet hosts and 561 larger-planet hosts in the mass-ratio-matched sample. The split uses the frozen catalog radius before the SOAR imaging outcome is applied.

The radius-split likelihood includes the angular resolved-target correction. Although both reported cutoffs end at 50 or 100 au, nearby pairs can cross the Gaia angular/contrast locus within those physical limits, so each radius group's target-specific design matrix includes that selection term. The resulting expectations and comparisons are reported below.

\subsubsection{Radius-split results}

\begin{figure}[t]
\centering
\includegraphics[width=\columnwidth]{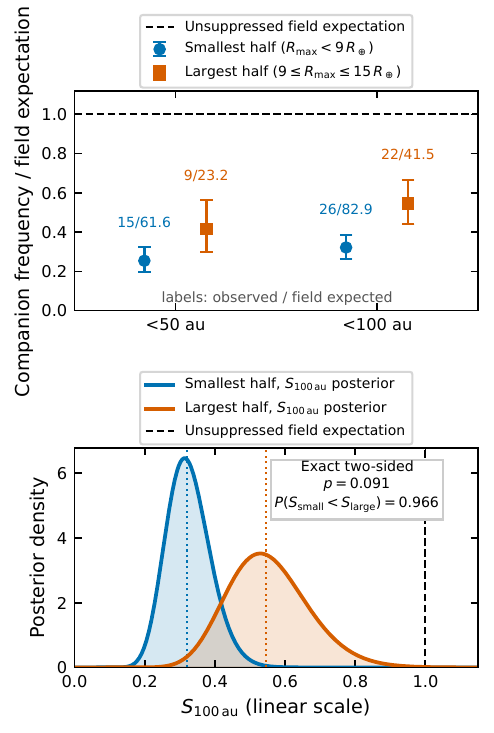}
\caption{Exploratory close-source frequencies in the two planet-radius groups. Top: posterior relative frequency at 50 and 100 au. Bottom: the two $S_{100\,{\rm au}}$ posterior densities, where $S_{100\,{\rm au}}$ denotes the relative frequency inside 100 au. Their overlap and the exact $p=0.091$ provide insufficient evidence for a radius dependence.}\label{fig:radiussplit}
\end{figure}

Both demographic subsamples show close-source deficits (Table~\ref{tab:radiussplit}; Figure~\ref{fig:radiussplit}). Among the 638 smaller-planet hosts, 15 sources are observed inside 50 au compared with 61.6 expected, and 26 are observed inside 100 au compared with 82.9 expected. The posterior relative frequencies are $S_{50\,{\rm au}}=0.254$ (0.196--0.323) and $S_{100\,{\rm au}}=0.322$ (0.264--0.388). Here $S_{50\,{\rm au}}$ and $S_{100\,{\rm au}}$ refer explicitly to relative source frequencies inside 50 and 100 au. Among the 561 larger-planet hosts, the corresponding results are 9/23.2 and 22/41.5, giving $S_{50\,{\rm au}}=0.417$ (0.298--0.564) and $S_{100\,{\rm au}}=0.546$ (0.440--0.667).

\begin{deluxetable*}{lrrrrr}
\tablecaption{Close-companion frequency by planet-radius group\label{tab:radiussplit}}
\tablehead{
\colhead{Largest planet/candidate}&\colhead{$N_\star$}&
\multicolumn{2}{c}{$s_{\rm proj}<50$ au}&
\multicolumn{2}{c}{$s_{\rm proj}<100$ au}\\
\colhead{}&\colhead{}&
\colhead{$N_{\rm obs}/N_{\rm field}$}&\colhead{$S$ (16--84\%)}&
\colhead{$N_{\rm obs}/N_{\rm field}$}&\colhead{$S$ (16--84\%)}
}
\startdata
$R_{\rm max}<9$ \rearth&638&15/61.6&$0.254\ (0.196$--$0.323)$&26/82.9&$0.322\ (0.264$--$0.388)$\\
$9\leq R_{\rm max}\leq15$ \rearth&561&9/23.2&$0.417\ (0.298$--$0.564)$&22/41.5&$0.546\ (0.440$--$0.667)$\\
\enddata
\tablecomments{Each host enters once, using its largest retained catalog radius. The target-specific field expectation accounts for the different distances, HRCam contrast curves, eccentricity and mass-ratio weights, and pair-by-pair angular resolved-target weighting. A conditional exact test of one common relative-frequency factor gives $p=0.259$ inside 50 au and $p=0.091$ inside 100 au.}
\end{deluxetable*}

The smaller-planet systems appear more depleted inside 100 au: $S_{\rm small}/S_{\rm large}=0.590$ with a 95\% interval of 0.337--1.040, and $P(S_{\rm small}<S_{\rm large})=0.966$. The latter is a posterior probability under the stated model and priors; the separate frequentist two-sided comparison gives $p=0.091$, and the 50 au comparison gives $p=0.259$. The evidence is too weak to conclude that the deficit depends on planet size.

\subsubsection{Interpretive cautions}

Three diagnostics caution against assigning the tentative trend to planet formation. First, it is concentrated among unconfirmed candidates: their post hoc 100 au comparison gives $p=0.036$, whereas the confirmed/known subset contains zero smaller-group and one larger-group detection. Second, the larger hosts are more distant and fainter: their median distance and TESS magnitude are 373 pc and 12.03, compared with 153 pc and 10.43 for the smaller group. The target-specific field model accounts for HRCam contrast sensitivity, but not for the radius-dependent probability that TESS detects and promotes a diluted transit. Third light preferentially removes shallow transits \citep{Lester2021,Bergsten2026}. Third, applying the primary-host dilution correction before dividing the sample moves seven systems above 9~\rearth, including three with a companion inside 50 au, but only systems with a detected companion receive that reclassification. We therefore retain catalog radii for the test and treat corrected radii only as a diagnostic.

The observed trend runs opposite to the simplest growth-limited interpretation. Shorter-lived, truncated disks should preferentially impede the assembly of massive cores and gaseous envelopes; \citet{Sullivan2024} accordingly found a deficit of sub-Neptunes relative to super-Earths in close Kepler binaries. That mechanism predicts stronger suppression toward larger planets, whereas the tentative difference here points below 9~\rearth. Our 9~\rearth\ division also separates broadly planet-sized candidates rather than resolving the 1--4~\rearth\ super-Earth/sub-Neptune structure. Both size groups show close-companion deficits, with the physical radius dependence unresolved.

The companion-mass-ratio split remains in Appendix~\ref{app:secondary}. Its fitted slope changes sign under plausible field-$q$ priors, leaving the suppression amplitude's mass-ratio dependence unresolved.

\subsection{Confirmed planets in candidate close binaries}\label{sec:survivors}

A population-level deficit can coexist with individual planets in close pairs. Confirmed planets in binaries below 100 au are already known outside the SOAR sample; Kepler-444A, for example, hosts five small planets while the BC pair lies at a projected 66 au \citep{Dupuy2016}. The few confirmed planets with SOAR-detected companions are complementary tests of survival in the binary-depleted regime, provided that the stellar association and transited component can be established.

Only two systems in the controlled sample combine a confirmed planet with a SOAR detection inside 100 au after literature vetting: HD~213885~b (TOI-141.01) and WASP-144~b (TOI-1928.01). The two-epoch SOAR astrometry establishes common motion for the HD~213885 companions. The 0.10 arcsec WASP-144 pair has a negligible local chance-alignment estimate, although a second epoch is still needed for a direct common-motion test and the planet-host component remains unknown.

HD 213885 b (TOI-141.01) is a 1.008-day, $1.74\pm0.05$ \rearth, $8.8\pm0.6$ $M_\oplus$ planet \citep{Espinoza2020}. On 2018 September 25, SOAR measured the inner companion at $\rho=0.4429$ arcsec, ${\rm PA}=239.5^\circ$, and $\Delta I=4.9$ mag, and the outer companion at 1.1999 arcsec, $305.2^\circ$, and 5.4 mag. On 2019 July 14 the respective measurements were 0.4450 arcsec, $238.9^\circ$, 4.8 mag and 1.2094 arcsec, $305.1^\circ$, 5.1 mag. Over 0.80 yr, both two-dimensional relative positions changed far less than expected for stationary background objects given the primary's 137 mas yr$^{-1}$ proper motion, establishing common motion. The Gaia local-density plug-in probability is numerically zero for both sources because no comparison source met the brightness cut. Using the latest contrasts, the companions contribute 2.1\% of the $I$-band flux; under the primary-host assumption the planet-radius factor is 1.011.

WASP-144 b (TOI-1928.01) is a 2.278-day planet with $M_p=0.44$ $M_{\rm Jup}$ and $R_p=0.85$ $R_{\rm Jup}$ \citep{Hellier2019}. SOAR detected an equal-brightness pair at 0.10 arcsec, or 36 au, in one epoch. Gaia leaves the pair unresolved, and the matched DR3 source has RUWE=4.89. Elevated RUWE supports unresolved astrometric structure, while the local-density chance-alignment estimate is numerically zero. A second epoch can test common motion; deriving an orbit at this scale will require a much longer baseline. If the catalog primary hosts the transit and the rounded $\Delta I=0$ value is accurate, dilution raises the planet radius by $\sqrt{2}$ to about 1.20 $R_{\rm Jup}$. Spatially resolved transit photometry or spectral disentangling of the blended system is needed to identify the planet-host component and revise the stellar and planet parameters.

TOI-2543.01, the third system identified in the preliminary catalog query, is a brown dwarf: \citet{Psaridi2022} measured $67.6\pm3.4$ $M_{\rm Jup}$. In the frozen 2026 July 30 TOI snapshot, its disposition column still reads CP even though the notes mark the signal as retired following the brown-dwarf mass measurement. The SOAR table contains a single-epoch 0.10 arcsec, $\Delta I=1.7$ detection, whereas the earlier ShARCS image in \citet{Psaridi2022} did not show a companion. A new resolved epoch is warranted, and the system is excluded from every planet-host statistic here.

Thus the close-source census supplies two valuable confirmed-planet systems: HD~213885 has common-motion companions, while WASP-144 still needs a second epoch and host-component identification.

\subsection{Signal dispositions and close visual detections}\label{sec:reliability}

The scarcity of confirmed planets in close pairs also raises a reliability question about the unresolved candidates. We performed a signal-level disposition audit among catalog signals whose host has a measured HRCam contrast curve, $3900\leq T_{\rm eff}\leq7400$ K, a known distance, and a catalog radius between 0.1 and 15~\rearth. Each signal was grouped by the nearest detected SOAR component within the fixed comparison domain $\Delta I\leq5.1$ and $1\leq s_{\rm proj}\leq5{,}000$ au. ``No detection'' therefore means no SOAR component in that stated domain, not that every possible companion has been excluded. CP and KP signals were counted as planets; FP and FA signals and the literature-vetted brown dwarfs were counted as known nonplanets; PC and APC signals remained unresolved. The audit retains rejected signals and is therefore broader than the 1,199-host demographic sample.

Inside 100 au, the audit contains 2 bona fide planet signals, 19 known nonplanets, and 58 unresolved candidates (Table~\ref{tab:reliability}). Among only the 21 classified outcomes, a Jeffreys binomial calculation gives a descriptive nonplanet fraction of 0.898 (0.822--0.950). The Jeffreys calculation is a binomial posterior using the standard weakly informative $\mathrm{Beta}(1/2,1/2)$ prior; it avoids zero-width or one-sided intervals for small samples. The corresponding nonplanet fractions are 0.568 (0.493--0.640) for signals whose nearest detection lies at 100--5,000 au and 0.427 (0.407--0.446) when no component in the comparison domain was detected. The close-detection nonplanet-to-planet odds exceed those of the wider group by a factor of 7.22 ($p=0.0096$) and those of the no-detection group by a factor of 12.78 ($p=1.34\times10^{-5}$), using two-sided Fisher exact tests. A Fisher test compares the four integer counts in a $2\times2$ table under the null hypothesis of no association and reports the probability of a table at least as imbalanced in either direction. The contrast is not solely a giant-candidate effect: below 9~\rearth, classified close systems contain 16 nonplanets and one planet, compared with 183 and 185 in the no-detection group ($p=2.3\times10^{-4}$).

\begin{deluxetable*}{lrrrrc}
\tablecaption{Disposition outcomes by nearest SOAR detection in the comparison domain\label{tab:reliability}}
\tablehead{\colhead{Nearest detection}&\colhead{$N_{\rm signal}$}&\colhead{Planet}&\colhead{Nonplanet}&\colhead{PC/APC}&\colhead{$f_{\rm nonplanet}$ (16--84\%)}}
\startdata
$<100$ au &79&2&19&58&$0.898\ (0.822$--$0.950)$\\
100--5,000 au &199&19&25&155&$0.568\ (0.493$--$0.640)$\\
No detection in comparison domain &1553&351&261&941&$0.427\ (0.407$--$0.446)$\\
\enddata
\tablecomments{The fixed comparison domain is $1\leq s_{\rm proj}\leq5{,}000$ au and $\Delta I\leq5.1$. ``Planet'' means CP/KP after the brown-dwarf literature audit; ``nonplanet'' means FP/FA or a vetted brown dwarf. The Jeffreys fraction describes only the classified outcomes; calibrating the reliability of an individual PC/APC requires the selection and diagnostic information discussed in the text.}
\end{deluxetable*}

These fractions show that already classified outcomes have a different composition in the close-detection group. The numerical 0.898 value describes that classified subset; extending it to unresolved candidates or individual false-positive probabilities would require additional calibration. Suspected eclipsing binaries and crowded targets were preferentially sent for high-resolution imaging, confirmed planets receive unequal follow-up, and the imaged source may be unbound or unrelated to the transit. A calibrated probability requires transit-shape and centroid information, component colors and stellar densities, and explicit modeling of the follow-up selection function.

The disposition analysis therefore establishes a compositional difference among already classified signals. Calibrated reliabilities for the unresolved close-pair candidates await a model of the follow-up selection and transit diagnostics.

\subsection{Robustness and statistical scope}

Reasonable changes to the host sample and field normalization test the stability of the main deficit. The same checks delineate the inference supported by the imaging census from effects that require an end-to-end TESS selection model.

The deficit persists in every tested subset, but its amplitude depends on disposition. The confirmed/known sample contains one source inside 100 au where 37.4 are expected, giving $S_{100\,{\rm au}}=0.045$ (0.019--0.088), whereas the candidate-only sample gives $0.548$ (0.473--0.631). This contrast may combine residual false positives, preferential validation of suspicious candidates, and a real difference in planet-host multiplicity; one confirmed/known detection cannot separate those effects. Restricting the sample to 500 pc gives $S_{100\,{\rm au}}=0.405$ (0.350--0.466), and an FGK-only cut gives $0.424$ (0.363--0.491).

The field-model normalization carries systematic uncertainty. As a sensitivity test, we assigned it a 10\% log-normal uncertainty, comparable to the field-population systematic discussed by \citet{MoeKratter2021}. Marginalizing over that term, meaning averaging the inference over its assumed range while weighting each value by its probability, gives approximate 68\% intervals of 0.232--0.360 at 50 au and 0.328--0.463 at 100 au. Uncertainty in the mass--contrast relation and TOI selection remains separate, while a plausible normalization shift leaves the deficit intact.

Transit dilution and catalog selection can account for part of the factor of 2.6--3.5 all-host deficit. This study stops short of an end-to-end TESS injection test matched to the cumulative target list and reports the nearby-source frequency among cataloged TOI hosts relative to a specified physical field-binary model. Intrinsic planet occurrence requires the additional completeness analysis outlined above.

\section{Discussion}\label{sec:discussion}

The statistical results point toward suppressed planet yields in close binaries, but that physical reading depends on association, deblending, and transit completeness. We first summarize the secure companion-frequency result, then examine the interpretations allowed by the present data and the observations needed to distinguish them.

\subsection{Interpretation of the cumulative survey}

The robust result is a deficit of nearby sources in the mass-matched TOI sample: $S_{50\,{\rm au}}=0.291$ inside 50 au and $S_{100\,{\rm au}}=0.391$ inside 100 au. The independent, calibrated 100--300 au annulus rises to $S=0.851$. A smooth semimajor-axis recovery law gives a midpoint of 74 au; the fitted step has $\Delta{\rm AIC}=1.7$, so the transition shape is not distinguished, and the midpoint remains conditional on the chosen functional form and outer field-rate asymptote. The pre-2021 cohort retains the close deficit, although its 28 matched sources leave the recovery shape weakly constrained.

The combined SOAR--Gaia comparison helps clarify what causes that recovery. When the Gaia common-motion catalog is used as the working wide survey, a multiplicity-conserving outward shift cannot account for the missing close companions. The Gaia portion contains 88 companions compared with 69.6 in the pairwise-corrected field model. This excess requires field-normalization and Gaia-selection uncertainties for broad compatibility and does not support a claim of exact agreement. The freely shifted fixed-multiplicity fit drives its center outward to the optimizer boundary and remains worse than recovery by $\Delta{\rm AIC}=16.5$. The literature-motivated 100 au distribution is worse by 60.2. Reducing its amplitude to 0.683 still leaves $\Delta{\rm AIC}=35.9$ and makes it a deficit model rather than pure redistribution. The combined result therefore indicates a close deficit followed by recovery toward a field-compatible wide population, not a deficit extending to 5,000 au.

Under the deliberately primary-only convention of Section~\ref{sec:occurrence}, the smooth law removes 17\% of the volume-limited system yield within the measured $q\geq0.4$ domain and 23\% after an all-$q$ extrapolation. These are illustrations, not per-star occurrence measurements. Because either binary component can supply a TOI, the per-star suppression required to produce the observed companion deficit is stronger than those system-level numbers imply. The M-dwarf sample shows a similar deficit using a different field model, but the small sample does not allow a precise comparison with solar-type stars.

\subsection{What the field comparison measures}

The Raghavan stars provide a field multiplicity reference rather than a planet-free control sample. Planet occurrence surveys imply that many of those systems, plausibly most, host planets despite the absence of planet-based selection \citep{MoeKratter2021}. The measured $S$ is therefore the relative frequency of a specified stellar-companion population among stars that entered a catalog through a detected transiting, predominantly close-in signal. It is not automatically the ratio of planet occurrence per star. A deficit could mean that close binaries suppress planets altogether, suppress the short-period planets to which TESS is most sensitive, or make those transits harder to detect and validate. Conversely, the opportunity for either component to supply the signal raises the probability that a binary enters the TOI sample and therefore acts against the observed deficit. Because the present forward model does not include this two-host enhancement, the measured close-binary suppression is likely conservative: a complete selection model would generally predict more binaries in the planet-selected sample before applying any physical suppression.

The distribution of planetary periods makes that distinction concrete. Among 1,284 retained signals in the demographic hosts with reported periods, the median is 4.89 days and 90.3\% are below 20 days. As an exploratory host-level test, we classified each host by its longest retained planet or candidate period and split 1,181 hosts at $P_{\rm max}=4.803$ days. Inside 50 au the inferred frequencies are $S=0.256$ and 0.297, with a conditional two-sided $p=0.829$. Inside 100 au the shorter-period group is instead less depleted, $S=0.499$ versus 0.301, but the two-sided comparison gives $p=0.101$. This post hoc split, chosen after examining the data and entangled with other differences between the groups, provides no evidence that the close-companion deficit is driven specifically by disruption of the shortest-period TESS planets.

The analogous close-binary result for Kepler hosts argues against an explanation unique to TESS's shorter observing windows \citep{Kraus2016}. Neither survey, however, isolates planets on multi-year orbits. A direct test would compare companion frequencies for well-characterized planets on short, intermediate, and multi-year orbits.

\subsection{Physical interpretation}

The local-density calculation shows that chance alignment is negligible for most close SOAR detections, while it becomes increasingly relevant for faint sources near an arcsecond and depends strongly on the target field. For physically associated companions, the deficit occurs where the companion can strongly alter a circumstellar disk. Binary torques truncate the disk, and shorter viscous timescales then accelerate gas loss \citep{ArtymowiczLubow1994,Cieza2009,Kraus2012}. ALMA observations of the 22 au FO Tau binary resolve compact disks consistent with tidal truncation and show that favorable, aligned systems can retain planet-forming material \citep{Tofflemire2024}. Common-motion confirmation and characterization of the hierarchy are still required for individual systems whose association is not already established.

Truncation also removes the outer solid reservoir. Dust-evolution models predict faster radial drift in compact binary disks \citep{Zagaria2021,Zagaria2023}. The global calculations of \citet{Venturini2026} connect that loss directly to reduced pebble-fed core growth and predict recovery over a broad separation range. Binary-driven eccentricity can increase planetesimal collision speeds, although disk gravity and apsidal alignment can preserve local regions of growth \citep{SilsbeeRafikov2021}. These formation-stage mechanisms naturally reduce the number of planets without requiring every surviving system to be dynamically unstable.

The M-dwarf result is qualitatively compatible with the same mechanism, but its Winters-based field model and small counts limit the comparison to that qualitative level.

Post-formation dynamics provide a second pathway. Secular perturbations and Kozai--Lidov cycles can excite eccentricity and inclination, reduce transit multiplicity, destabilize planets, or drive high-eccentricity migration \citep{Marzari2019,MoeKratter2021}. To leading order, the Kozai--Lidov timescale depends only weakly on companion mass ratio across the range tested here. The fitted relation between suppression and $q$ has a slope consistent with zero and changes sign under alternate field priors, leaving disk truncation and dynamical forcing observationally degenerate.

\subsection{Planet properties and remaining biases}

Both planet-radius groups show deficits, but their difference remains weakly constrained. Because the 9~\rearth\ boundary lies well above the small-planet radius range, we cannot directly compare the nominally lower $S_{100\,{\rm au}}$ here with the super-Earth/sub-Neptune result of \citet{Sullivan2024}.

Selection provides a simpler explanation for the observed trend. The candidate-only comparison is stronger than the full-sample result, while the confirmed/known subsample has too few close detections to show any radius dependence. The larger-radius group is also more distant and fainter, and correcting radii only for detected companions creates an outcome-conditioned split. Unresolved companions additionally change the stellar catalog, target selection, and transit-detection efficiency. \citet{Bergsten2026} found that including these effects increases representative Kepler small-planet occurrence estimates by factors of about 1.08--1.19. An analogous TESS analysis must inject transits around both binary components, use sector-specific noise and apertures, include candidate reliability, and fit the planet and companion samples jointly.

The present result remains conditional on a solar-type field model. The matched empirical mass-ratio calculation improves on a fixed contrast, but it still assumes a dwarf mass--luminosity relation while holding metallicity, age, and unresolved multiplicity fixed. This limitation is most acute for the 66 demographic hosts assigned temperature-based masses. Many are close pairs whose TIC parameters come from blended photometry and can be strongly biased \citep{FurlanHowell2020}. Retaining them prevents a demonstrably multiplicity-dependent catalog cut, but their individual $q$ estimates and candidate interpretations require resolved characterization.

For the wide-separation comparison we adopted the explicit Gaia DR3 working selection in Section~\ref{sec:gaia}, rather than treating the common-motion catalog only as an association list. The simulation applies the same primary eligibility, $\rho\geq1$ arcsec, $G_2\leq20.7$, projected-separation, and empirical $q\geq0.4$ cuts as the observed catalog. The parallax and proper-motion criteria strongly reduce chance alignments, and any remaining contaminants raise the observed count rather than create a deficit. Source-level recovery for scanning law, crowding, color, and problematic astrometric solutions is omitted; the resulting incompleteness would lower the observed wide count.

Equations~\ref{eq:gaiaangular}--\ref{eq:widefactor} address two separate selection requirements: catalog-resolved pairs should not receive the brightness-selection advantage appropriate to unresolved systems, and the field mass-ratio law must vary with primary mass and separation. The resolved-target term removes each synthetic pair's own brightness boost in angular/contrast space. Independently, the El-Badry term assigns event-level mass-ratio weights as a function of primary mass and projected separation before HRCam or Gaia filtering. Gaia completeness and the heterogeneous routes by which TOIs entered the catalog remain additional sources of uncertainty. The resulting prediction is 69.6 Gaia companions, compared with 88 observed. A two-component planet-selection advantage could contribute to this wide excess, but its size changes with contrast, aperture, and vetting, so we do not use the excess as an empirical correction. A joint occurrence analysis should fit that selection and an outer companion-frequency normalization simultaneously rather than force a universal factor of two. We therefore describe the wide comparison as broadly compatible only after allowing for those systematic uncertainties, but the uncertainty is too large to show that the frequency has returned exactly to the field value. The multiplicity-conserving 100 au shift remains strongly disfavored across the same response. \textsc{paired} serves a different purpose: its sparse RVS coverage identifies candidate inner subsystems among apparently wide pairs rather than setting the wide imaging selection.

\subsection{Deblending and host-star identification}\label{sec:deblendingfuture}

The resolved contrasts make it possible to improve the stellar and planetary parameters of many close systems, but a reliable recalculation requires a joint two-star model rather than an additional dilution factor. Such a model would fit the blended Gaia, TIC, 2MASS, and WISE photometry and Gaia parallax simultaneously with the SOAR $\Delta I$ measurement, any resolved contrasts in other bands, and spectroscopic constraints. For a likely bound pair, a coeval, common-metallicity isochrone prior would connect the two components while allowing separate posterior distributions for $M_A$, $R_A$, $T_A$, $M_B$, $R_B$, and $T_B$. This approach has already been demonstrated by fitting unresolved photometry and spectroscopy together with resolved contrasts in Kepler planet-host binaries \citep{Sullivan2022}. A single SOAR contrast supports a decomposition under a bound-main-sequence assumption; background stars, evolved companions, and unresolved triples require separate hypotheses.

For a planet assigned to the primary, the self-consistent correction would be
\begin{equation}
R_{p,A}=R_{p,{\rm TIC}}
\left(\frac{R_{A,{\rm deblend}}}{R_{\rm TIC}}\right)X_{R,A},
\label{eq:deblendplanet}
\end{equation}
where the first factor revises the stellar radius and $X_{R,A}$ corrects the transit dilution. These terms can reinforce or partially cancel one another. A secondary-host solution must instead use the deblended $R_B$ and the secondary's bandpass-specific flux fraction. The radius corrections in Section~\ref{sec:radius} account only for dilution; final planet radii require the full stellar and host-star analysis.

Identifying which component is transited is a separate inference. The strongest evidence would come from spatially resolved in-transit photometry or component-resolved spectroscopy and radial velocities. Multicolor transit depths, difference-image centroids for the wider pairs, and transit-shape diagnostics can reject additional blend scenarios. Asterodensity profiling provides a scalable route for the unresolved pairs: the transit duration and shape, together with assumptions about the orbit, imply a stellar-density distribution, which is compared with the independently deblended densities of A and B under the two possible host assignments. Recent Bayesian applications show both the promise and the limitation of this method: some systems receive strong circumprimary assignments, while many remain ambiguous because of impact-parameter, eccentricity, signal-to-noise, and nearly equal-density degeneracies \citep{BurnsWatson2026}.

In this notation, A is the catalog primary, B is the resolved secondary, and $D$ denotes all data used in the host comparison, such as the transit light curve, resolved contrasts, deblended stellar properties, and any centroid or spectroscopic constraints. The vertical bar is read as ``given.'' Thus $P(A\mid D)$ is the posterior probability that A is transited given the data $D$, while $P(B\mid D)$ is the corresponding probability that B is transited. A useful catalog would report both values together with posterior probabilities for eclipsing-binary and background-eclipsing-binary alternatives, rather than forcing a host label for every TOI. If those alternatives form an exhaustive set, their probabilities and the A- and B-host probabilities sum to unity. Priors based on the planet-radius distribution would require particular care, because using the inferred radius distribution to assign hosts and then using those assignments to measure the same distribution could bias the result.

A future catalog-scale analysis could combine joint two-star stellar posteriors with primary- and secondary-host planet-radius posteriors and probabilistic host and false-positive classifications. Marginalizing over those uncertainties would permit an improved planet-radius distribution and suppression analysis. Those calculations require data and assumptions beyond the imaging census considered here.

\subsection{Specific next tests}

Several tests can be performed with this release or modest follow-up. Multi-epoch astrometry for the 48 demographic hosts with detections inside 100 au would separate background objects from common-motion pairs and, for the closest or fastest systems, detect relative acceleration. At separations near 50 au the periods are typically centuries, so short arcs cannot deliver useful visual orbits. A much longer baseline and external radial-velocity or absolute-astrometric constraints would eventually permit semimajor-axis inference for selected systems.

Two-color speckle photometry would supply key additional constraints for the deblending analysis in Section~\ref{sec:deblendingfuture}. Equal-brightness candidates are especially informative because the field twin excess is separation dependent and drives much of the prior sensitivity \citep{MoeDiStefano2017}. For the 0.10 arcsec WASP-144 pair, spatially resolved transit photometry or analysis of the blended spectral line profiles is more realistic than direct component-resolved high-resolution spectroscopy.

The ten \textsc{paired}-only systems in the jointly covered sample, especially the seven with a resolved SOAR component beyond 300 au, are efficient hierarchical-system targets. Multi-epoch spectroscopy can test whether their excess Gaia RVS noise comes from an inner stellar subsystem. A confirmed inner subsystem plus the imaged wide companion would make the target at least triple. The TESS signal could still orbit the wide imaged star, so the host component must be established before deciding which orbit truncated the planet-forming disk; assigning the planet to an unseen RV subsystem without transit-localization evidence would be unwarranted.

An end-to-end TESS injection--recovery analysis can distinguish a physical binary effect from dilution and catalog selection and determine whether the tentative radius difference survives. It should inject planets around both components, use sector-specific apertures and noise, preserve the binary-dependent stellar catalog, and assign each PC/APC a reliability weight.

A confirmed-planet sample large enough to divide the 1--4~\rearth\ population into super-Earth and sub-Neptune bins would provide a direct TESS comparison with \citet{Sullivan2024}. Planet multiplicity and mutual inclination are complementary outcomes: post-formation forcing predicts fewer transiting multis even where planets still form. These tests are more informative than interpreting the present post hoc candidate-only contrast as a formation signal.

\section{Summary}\label{sec:summary}

We analyzed the complete SOAR TESS Survey as one cumulative program. The main results are:
\begin{enumerate}
\setlength{\itemsep}{0.45\baselineskip}
\setlength{\parsep}{0pt}
\item HRCam observed 2,982 unique TIC targets. The latest-epoch catalog contains 731 source detections around 689 hosts, and the post-Paper-II table publishes 433 measurements around 405 hosts. Ordinary new targets were selected by TOI status, brightness, sky position, and observability without multiplicity information; known binaries were placed on a separate repeat-observation list and could not add new demographic systems.
\item The primary sample contains 1,199 dwarf primaries and 162 visual companions with empirical $q\geq0.4$. The same Pecaut--Mamajek relation defines observed $q$, simulated contrast, detectability, and magnitude weighting. A $\Delta I\leq5.1$ robustness check changes the observed counts only from 24 to 25 inside 50 au and from 48 to 51 inside 100 au.
\item We observe 24 sources inside 50 au compared with 84.8 field binaries and 48 inside 100 au compared with 124.4. The corresponding relative frequencies are $S_{50\,{\rm au}}=0.291^{+0.062}_{-0.054}$ inside 50 au and $S_{100\,{\rm au}}=0.391^{+0.058}_{-0.053}$ inside 100 au. The independently calibrated 100--300 au value is $S=0.851$ (0.731--0.984).
\item The field model uses a truncated-Gaussian eccentricity distribution and the mass- and separation-dependent El-Badry $q$ distribution event by event. Resolved-target selection is likewise imposed pair by pair in angular-separation--contrast space. A continuous HRCam-only model gives $a_{50}=74$ au (57--101 au), while the step has comparable AIC ($\Delta{\rm AIC}=1.7$). In the 1--5,000 au comparison, the Gaia range contains 88 companions compared with 69.6 in the pairwise-weighted field model. A multiplicity-conserving shift remains strongly disfavored, including the fixed 100 au model ($\Delta{\rm AIC}=60.2$).
\item Under a primary-only, per-system convention, the fitted law removes 17.0\% (15.5--18.3\%) of volume-limited yield over $q\geq0.4$ and 23.3\% (21.2--25.1\%) under an all-$q$ extrapolation. These are illustrative system-yield reductions, not per-star occurrence rates. Because either component can supply a TOI, accounting for secondary-host detections makes the inferred suppression per star stronger. The planet-radius groups differ with $p=0.091$ inside 100 au. The fiducial mass-ratio slope is consistent with zero, and alternate field-$q$ priors change its sign.
\item The 89-host M-dwarf experiment also shows a deficit relative to its Winters-based field model: 2 sources are observed inside 50 au where 16.7 are expected. A $q\geq0.4$ comparison gives no significant amplitude difference from the solar-type result, while the distinct field models limit the inference to a qualitative cross-mass comparison.
\item Local Gaia source densities quantify association probabilistically for the post-Paper-II table; close chance alignments are generally negligible, while faint sources near an arcsecond require target-specific assessment. HD~213885's two companions show common motion. Among signals with a SOAR detection inside 100 au, 19 are known nonplanets and 58 remain PC/APC; the classified-outcome fraction is descriptive rather than an individual reliability estimate. Gaia identifies 510 common-motion companions around 451 hosts; 88 enter the matched 300--5,000 au wide-survey comparison, while RUWE and \textsc{paired} remain complementary unresolved-binary diagnostics.
\end{enumerate}

The cumulative survey establishes a robust close-companion deficit among selected TOIs relative to the adopted, contrast-limit-convolved field-binary model. Interpreting that deficit as lower planet occurrence in physical binaries is plausible but conditional on association probabilities, two-component TOI selection, transit completeness, and candidate reliability.

\section*{Data Availability}
The electronic submission includes the machine-readable SOAR and Gaia companion catalogs, target-level HRCam contrast limits, and field mass-ratio parameters. They will be freely available with the published article.\footnote{During peer review, the machine-readable tables may be requested from the corresponding author.}

\acknowledgments
We thank the SOAR and CTIO staff and the TESS Follow-up Observing Program community. The SOAR Telescope is a joint project of the Brazilian Minist\'erio da Ci\^encia, Tecnologia e Inova\c{c}\~oes, NSF's NOIRLab, the University of North Carolina at Chapel Hill, and Michigan State University. This research made use of ExoFOP-TESS, operated by the California Institute of Technology under contract with NASA through the Exoplanet Exploration Program \citep{ExoFOPTESS}, and the VizieR catalog access tool at CDS, Strasbourg, France. This work has made use of data from the European Space Agency (ESA) mission \textit{Gaia}, processed by the Gaia Data Processing and Analysis Consortium (DPAC). Funding for DPAC has been provided by national institutions, in particular the institutions participating in the Gaia Multilateral Agreement. This work made use of NumPy, pandas, Matplotlib, Astropy, and \textsc{paired} \citep{Harris2020,pandas,Hunter2007,Astropy2013,Astropy2018,Astropy2022,ChancePAIRED}.

\facility{TESS, SOAR (HRCam), Gaia}
\software{Astropy, NumPy, pandas, Matplotlib, PAIRED}

\clearpage
\appendix
\renewcommand{\theHequation}{appendix.\Alph{section}.\arabic{equation}}
\onecolumngrid
\setlength{\textfloatsep}{7pt plus 2pt minus 2pt}
\setlength{\floatsep}{7pt plus 2pt minus 2pt}
\setlength{\intextsep}{7pt plus 2pt minus 2pt}

\input{appendix_context}
\input{appendix_secondary}
\input{appendix_model_fits}

\section{Companion Catalogs and Autocorrelation Functions}\label{app:newcompanions}

\subsection{New SOAR/HRCam Companion Measurements}

Table~\ref{tab:newmeasurements} shows representative rows from the 433 component measurements obtained after 2020 December 31; the complete catalog is supplied in machine-readable form. The measurements come from 429 unique archive rows around 405 hosts, with one row per component and epoch. Repeated epochs are retained as separate measurements, and position angles are available for all 433 entries. We adopt the B-row astrometry from the nightly HRCam reduction logs: separations retain four decimal places, while position angle and $\Delta I$ retain the reported precision. The table includes the wide TOI-5831 source at $\rho=4.7723$ arcsec, ${\rm PA}=294.5^\circ$, and $\Delta I=0.7$ mag; the 2024 January 8 TOI-6094 source at $\rho=0.0378$ arcsec, ${\rm PA}=81.2^\circ$, and $\Delta I=0.7$ mag; and the TOI-6461 hierarchy, with A--BC and an inner B--C pair at 0.2616 arcsec, $281.1^\circ$, and 0.5 mag.

The two TOI-4182 epochs show why measurements are not collapsed: $\rho=0.04$ arcsec and $\Delta I=0.2$ mag on 2021 November 20, followed by $\rho=0.0452$ arcsec, ${\rm PA}=223.2^\circ$, and a rounded $\Delta I=0.0$ mag on 2024 January 8. Both are reproduced below. The occurrence calculation uses the latest epoch by its predeclared rule; the publication table preserves both observations.

Table~\ref{tab:newmeasurements}, the target-level HRCam contrast limits, and the field mass-ratio parameters are supplied in machine-readable form. The contrast-limit product gives TIC, angular separation, and $5\sigma$ $\Delta I$ at every interpolated curve point for all 2,684 selected targets, including the 2024 January nondetections. In Table~\ref{tab:newmeasurements}, ``Pair'' distinguishes A--BC from inner B--C or Ba--Bb measurements, and the astrometry retains the precision of the source reduction. The printed and machine-readable versions both report $P_{\rm bg}$, the local Gaia chance-alignment probability defined by Equation~\ref{eq:pbg}; the machine-readable table additionally provides the local-density inputs.

\startlongtable
\input{sample_new_companion_measurements.tex}

\subsection{Autocorrelation Functions}

Figures~\ref{fig:newacf1}--\ref{fig:newacf12} show calibrated SOAR/HRCam autocorrelation functions for all 405 hosts in Table~\ref{tab:newmeasurements}. The TOI-5831 panel uses a wider scale to show its 4.7723 arcsec source, while the 0.0378 arcsec TOI-6094 source appears as a central elongation. When several products exist, we show the epoch with the smallest PSF FWHM. Each panel uses an inverse linear scale that suppresses processing structure without hiding the companion peak. Arrows stop short of one point-symmetric peak and do not encode absolute quadrant; north is up and east is left.

\begin{figure*}[p]
\centering
\includegraphics[width=0.93\textwidth]{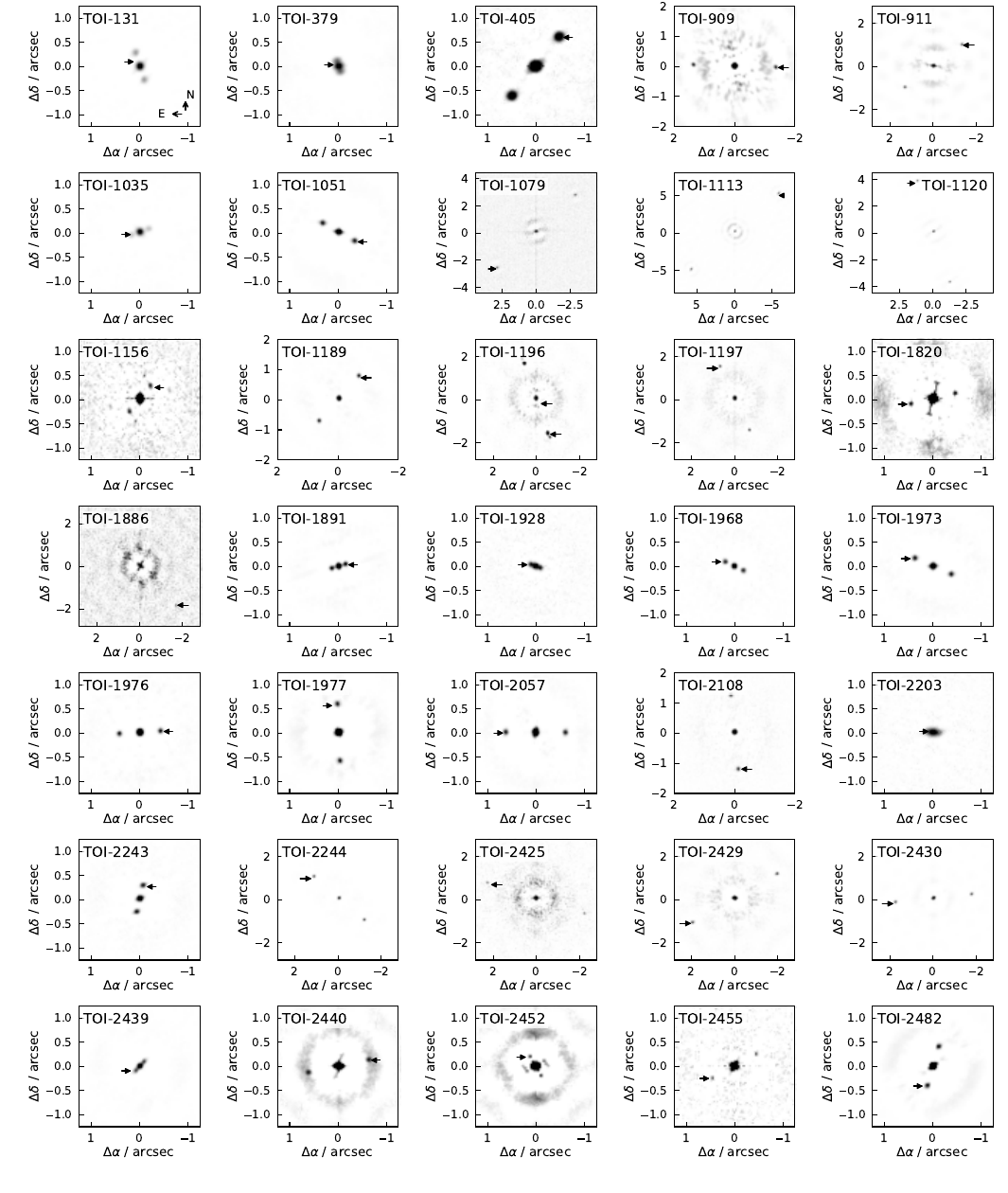}
\caption{SOAR/HRCam autocorrelation functions for TOI-131 through TOI-2482, ordered by TOI. Each panel uses an independently selected inverse linear scale; positive $\Delta\alpha$ is plotted to the left and positive $\Delta\delta$ upward. Each high-contrast outlined arrow stops just short of one point-symmetric companion peak. Absolute astrometry is given in Table~\ref{tab:newmeasurements}. The first panel includes the north--east compass.\label{fig:newacf1}}
\end{figure*}

\begin{figure*}[p]
\centering
\includegraphics[width=0.93\textwidth]{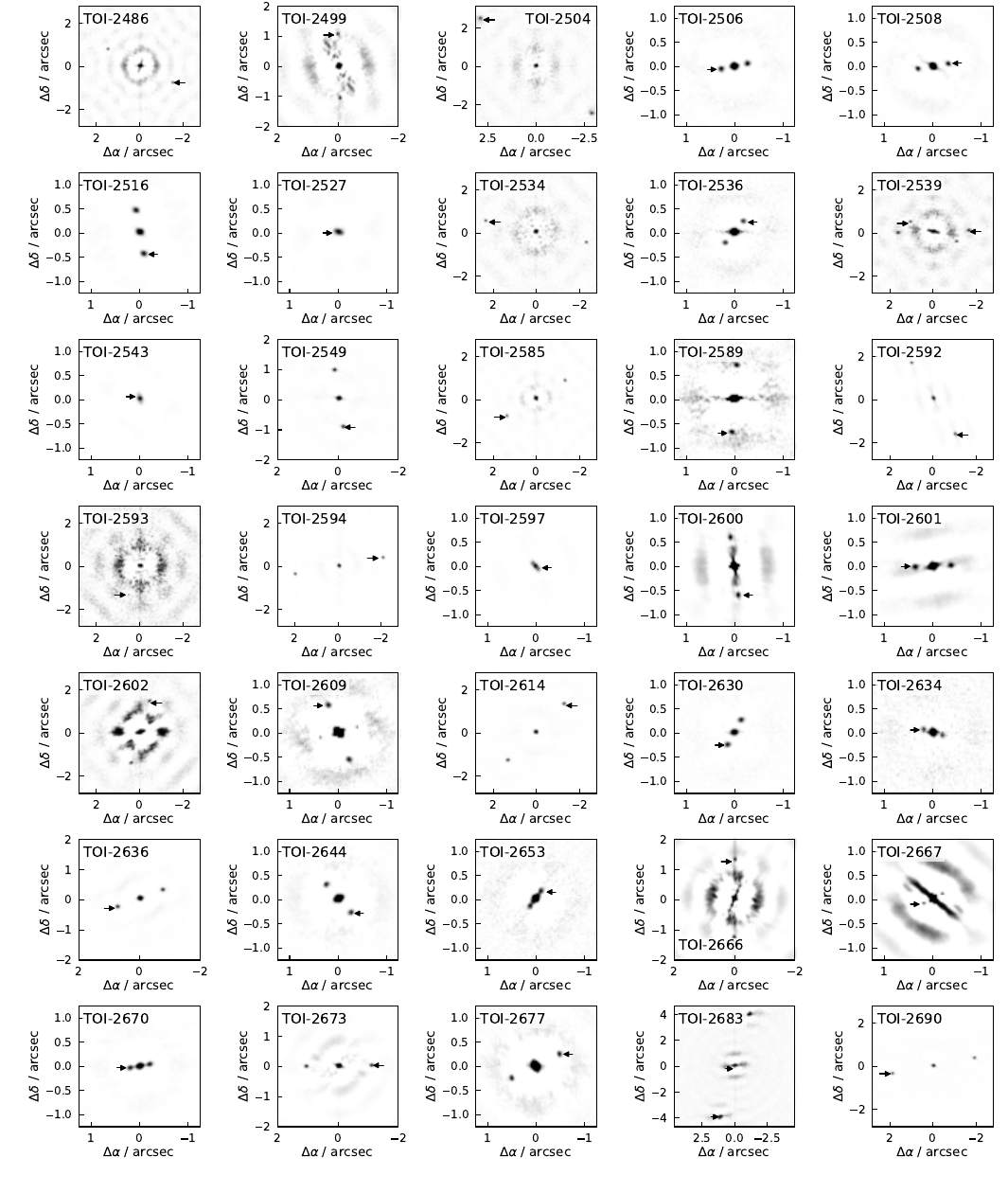}
\caption{Continuation of Figure~\ref{fig:newacf1} for TOI-2486 through TOI-2690.\label{fig:newacf2}}
\end{figure*}

\begin{figure*}[p]
\centering
\includegraphics[width=0.93\textwidth]{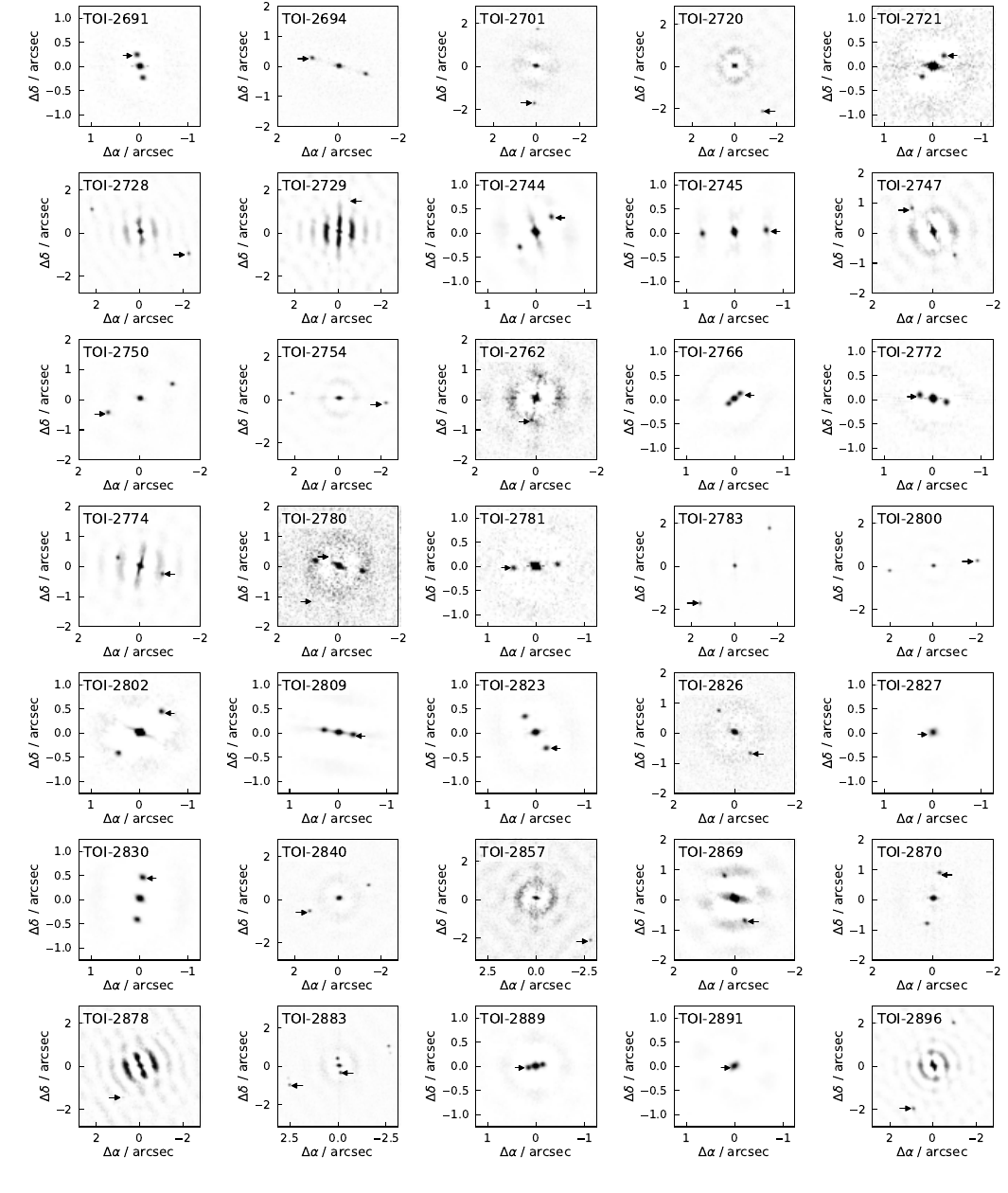}
\caption{Continuation of Figure~\ref{fig:newacf2} for TOI-2691 through TOI-2896.\label{fig:newacf3}}
\end{figure*}

\begin{figure*}[p]
\centering
\includegraphics[width=0.93\textwidth]{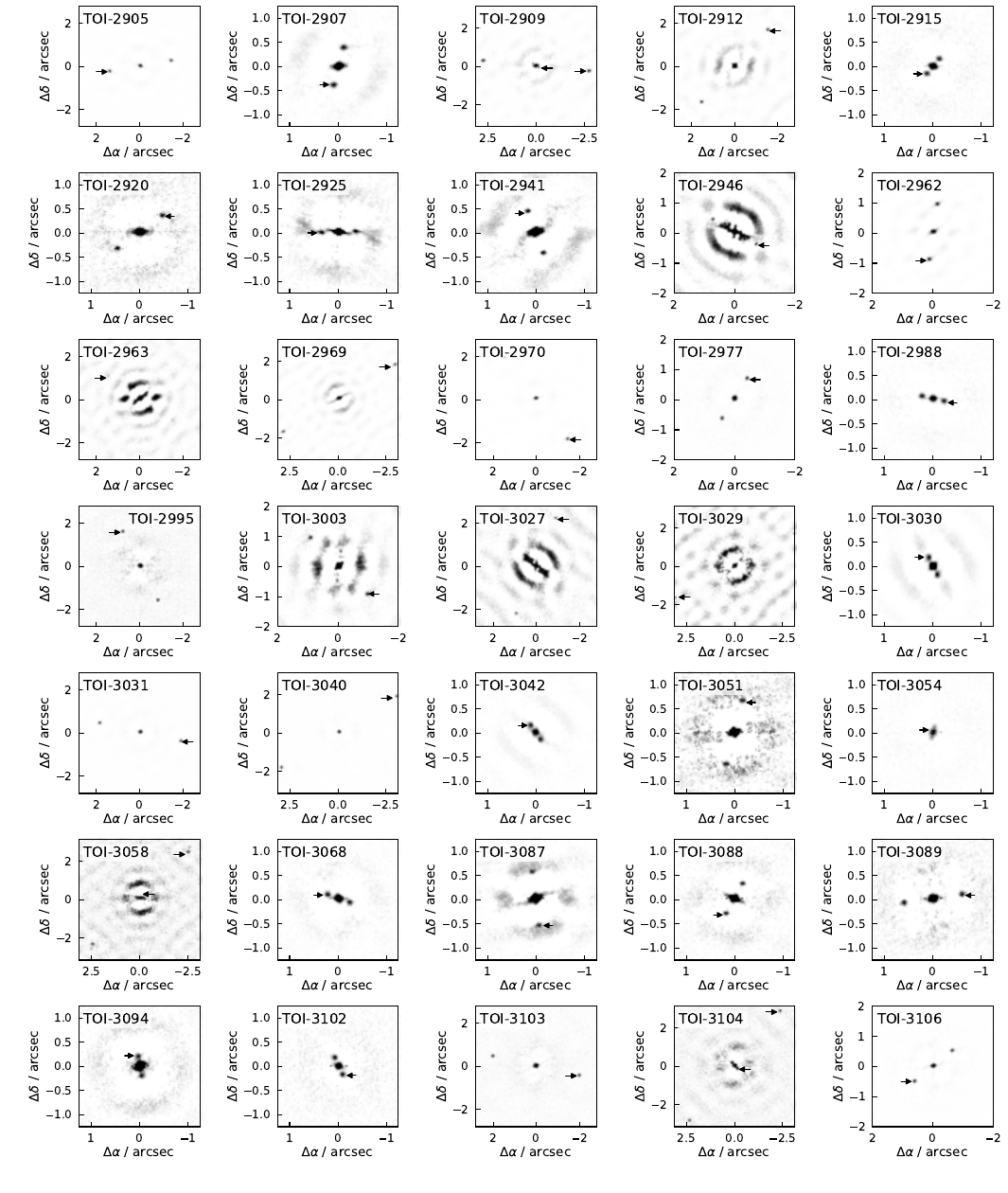}
\caption{Continuation of Figure~\ref{fig:newacf3} for TOI-2905 through TOI-3106.\label{fig:newacf4}}
\end{figure*}

\begin{figure*}[p]
\centering
\includegraphics[width=0.93\textwidth]{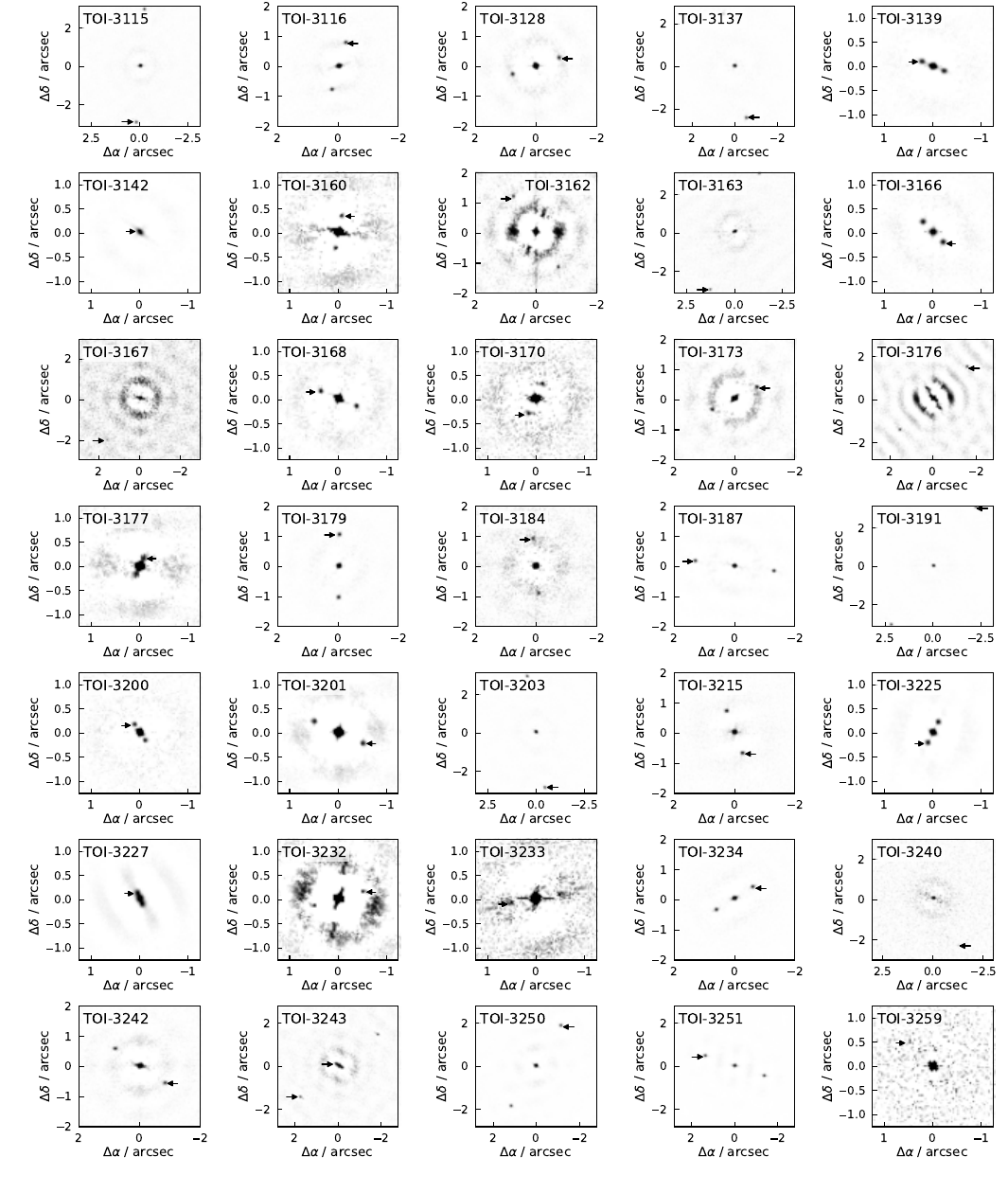}
\caption{Continuation of Figure~\ref{fig:newacf4} for TOI-3115 through TOI-3259.\label{fig:newacf5}}
\end{figure*}

\begin{figure*}[p]
\centering
\includegraphics[width=0.93\textwidth]{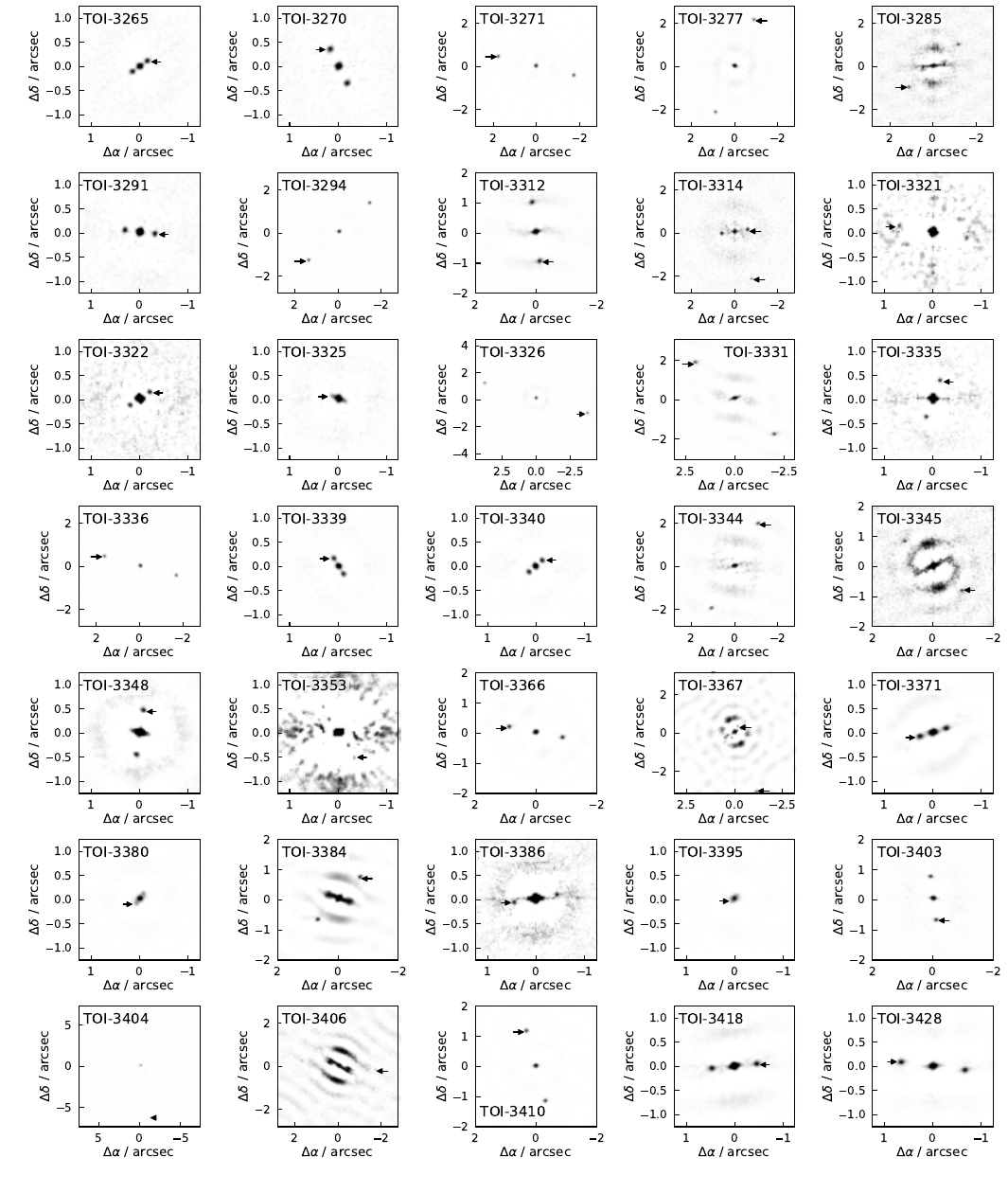}
\caption{Continuation of Figure~\ref{fig:newacf5} for TOI-3265 through TOI-3428.\label{fig:newacf6}}
\end{figure*}

\begin{figure*}[p]
\centering
\includegraphics[width=0.93\textwidth]{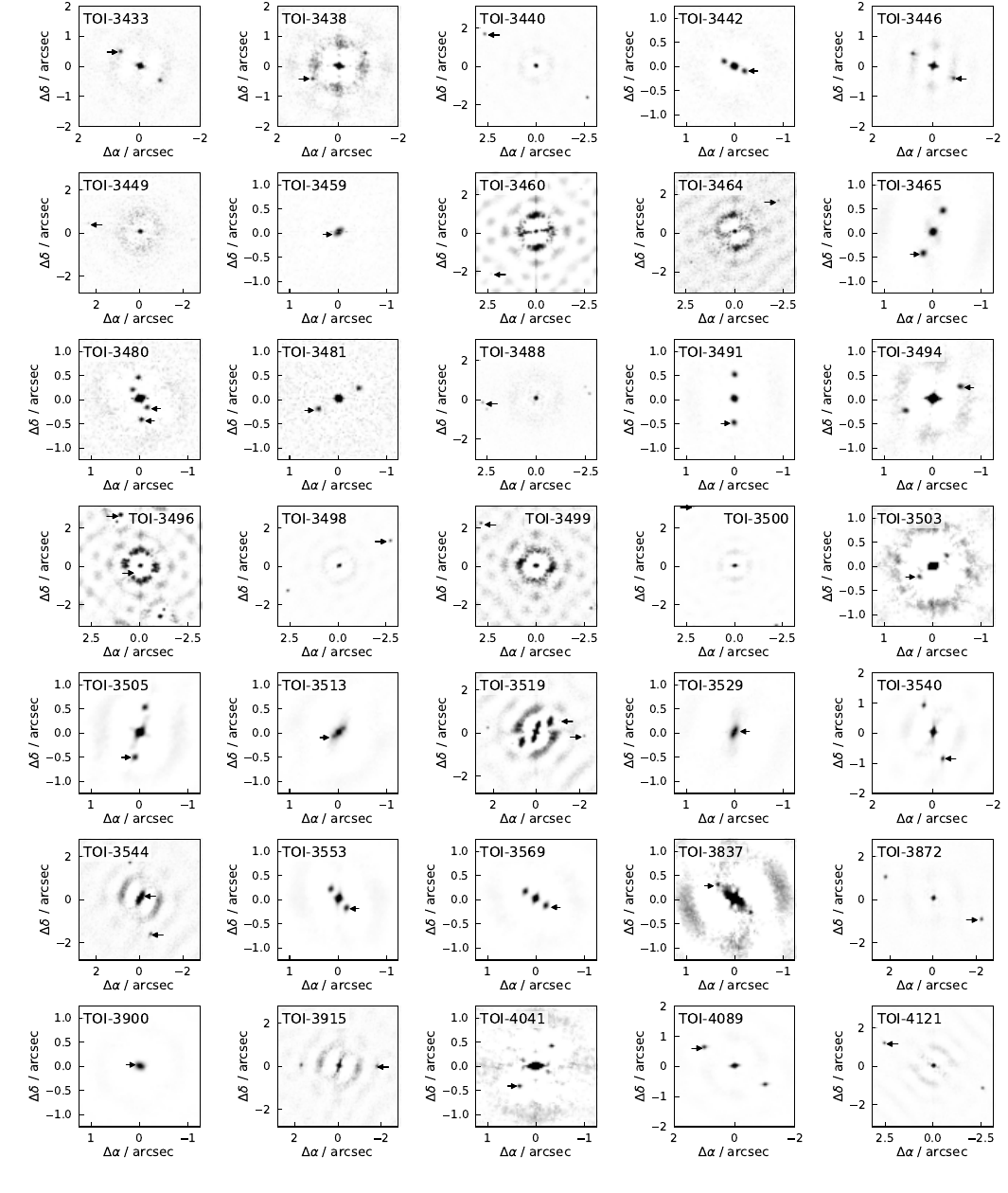}
\caption{Continuation of Figure~\ref{fig:newacf6} for TOI-3433 through TOI-4121.\label{fig:newacf7}}
\end{figure*}

\begin{figure*}[p]
\centering
\includegraphics[width=0.93\textwidth]{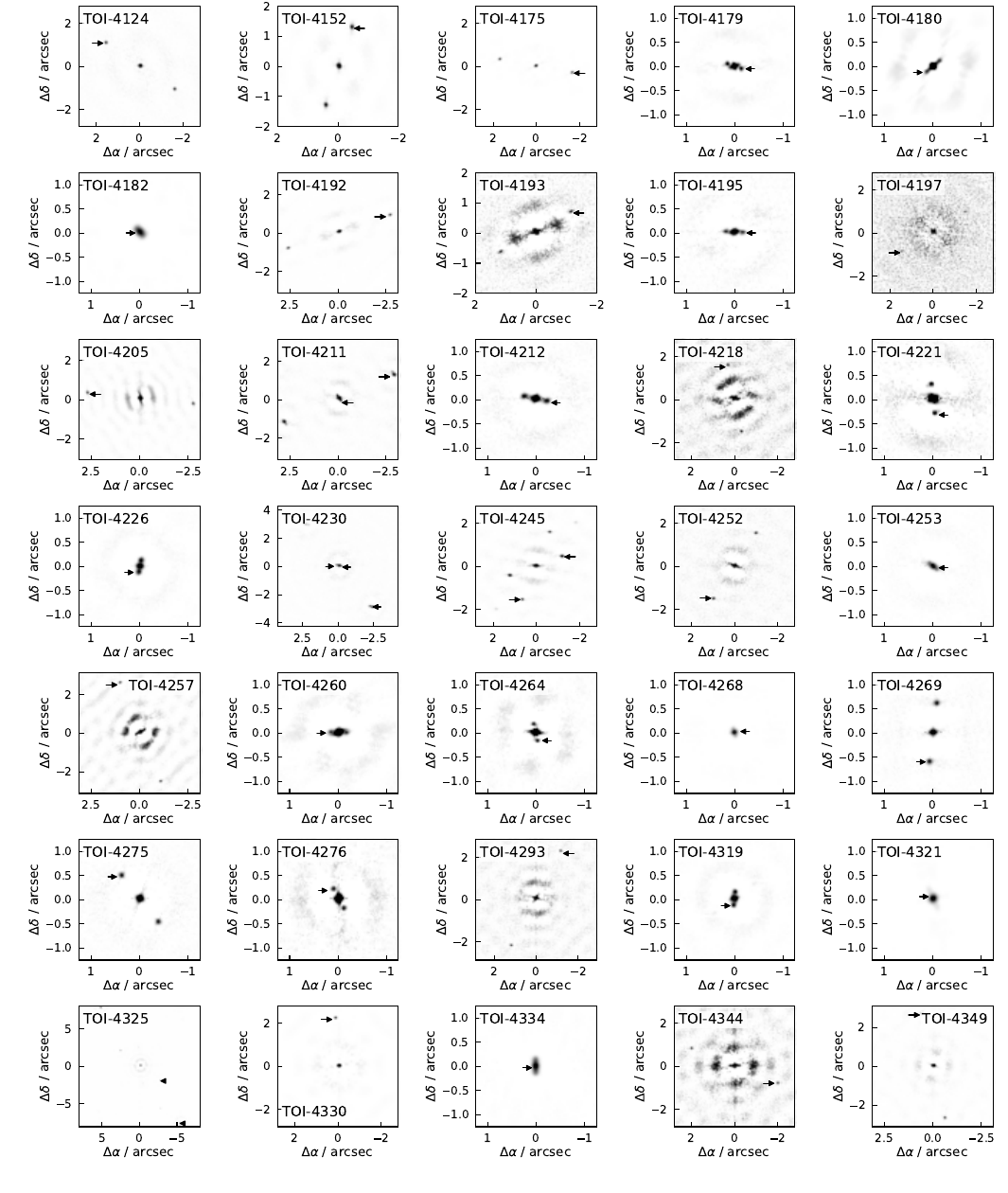}
\caption{Continuation of Figure~\ref{fig:newacf7} for TOI-4124 through TOI-4349.\label{fig:newacf8}}
\end{figure*}

\begin{figure*}[p]
\centering
\includegraphics[width=0.93\textwidth]{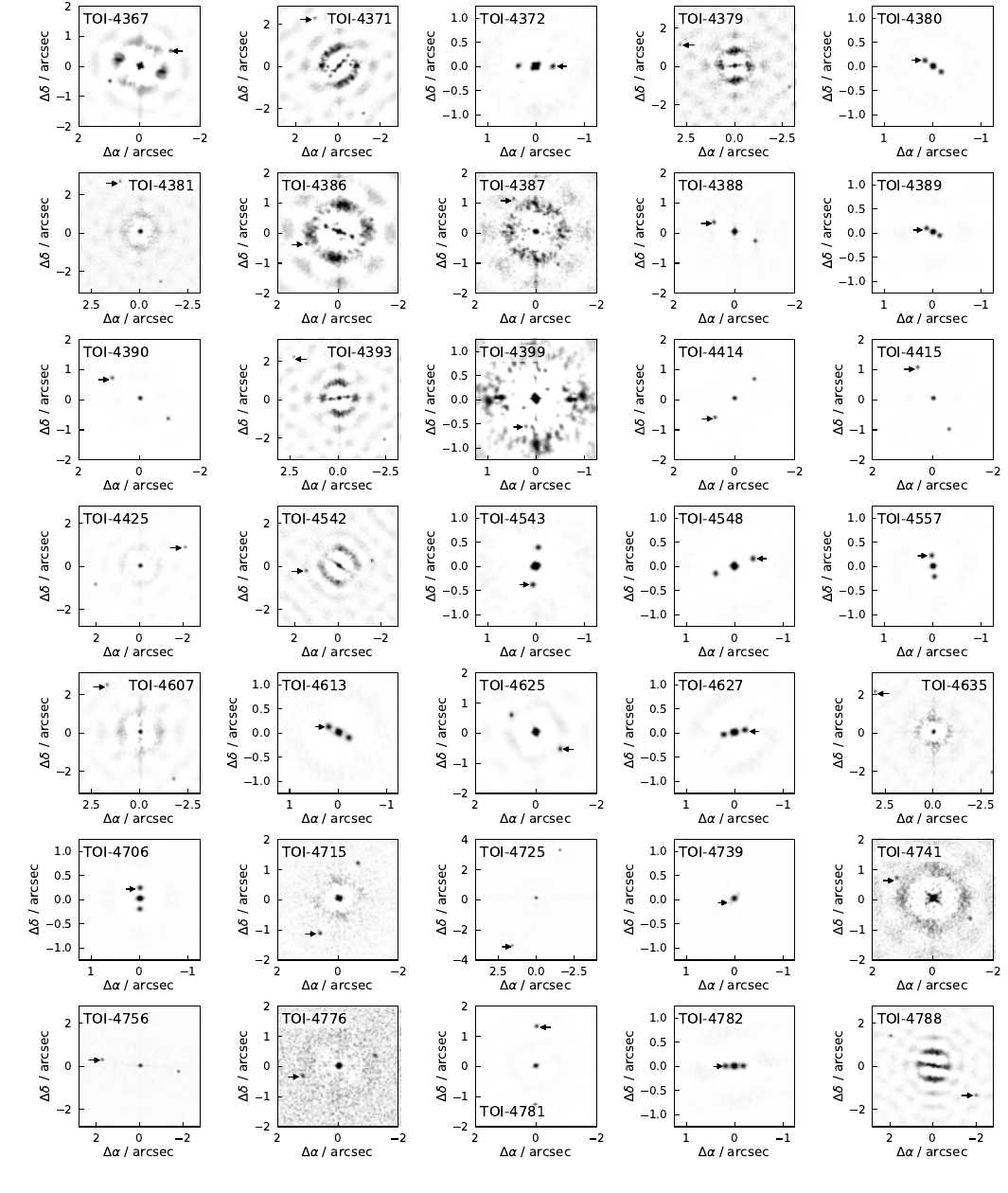}
\caption{Continuation of Figure~\ref{fig:newacf8} for TOI-4367 through TOI-4788.\label{fig:newacf9}}
\end{figure*}

\begin{figure*}[p]
\centering
\includegraphics[width=0.93\textwidth]{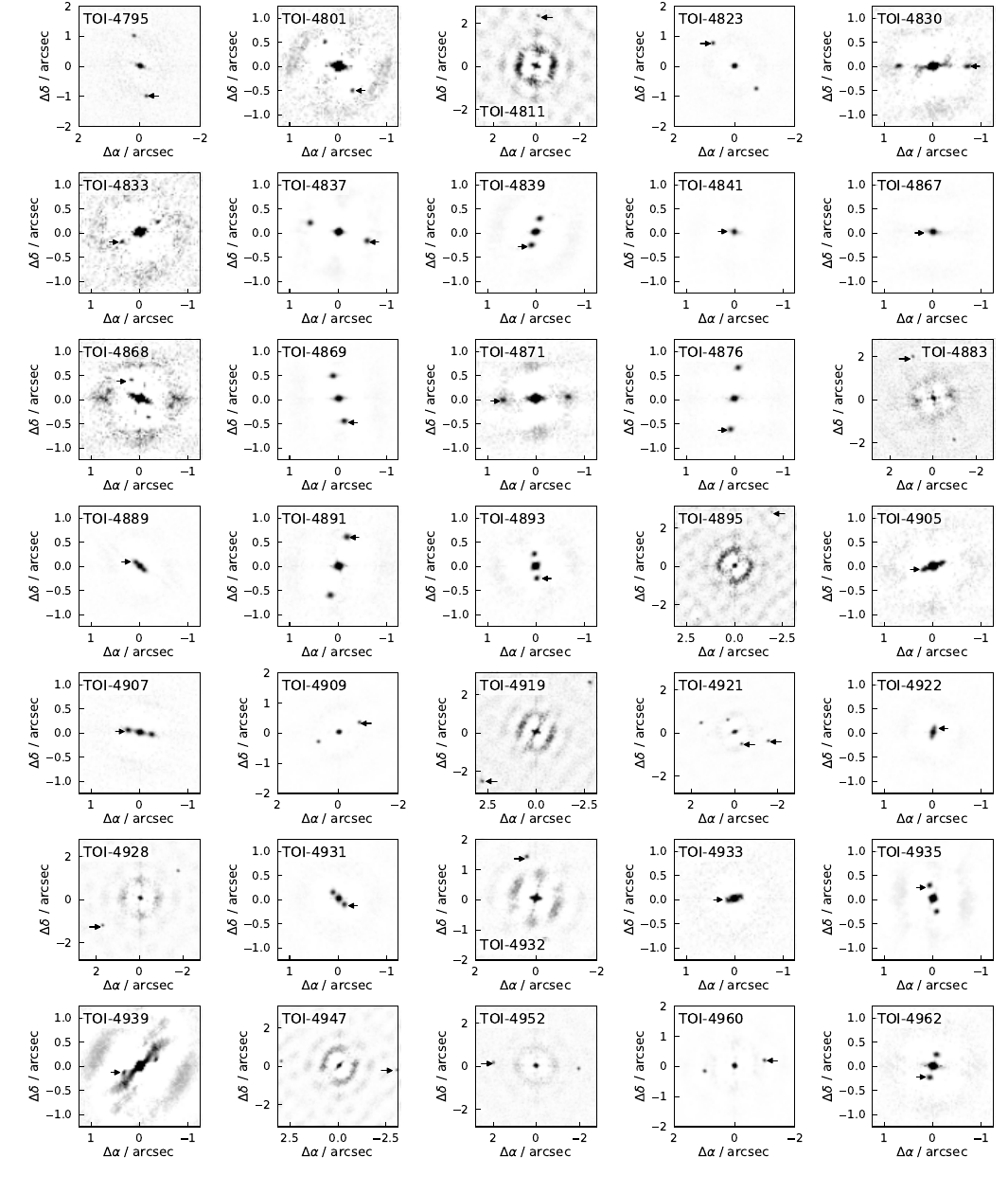}
\caption{Continuation of Figure~\ref{fig:newacf9} for TOI-4795 through TOI-4962.\label{fig:newacf10}}
\end{figure*}

\begin{figure*}[p]
\centering
\includegraphics[width=0.93\textwidth]{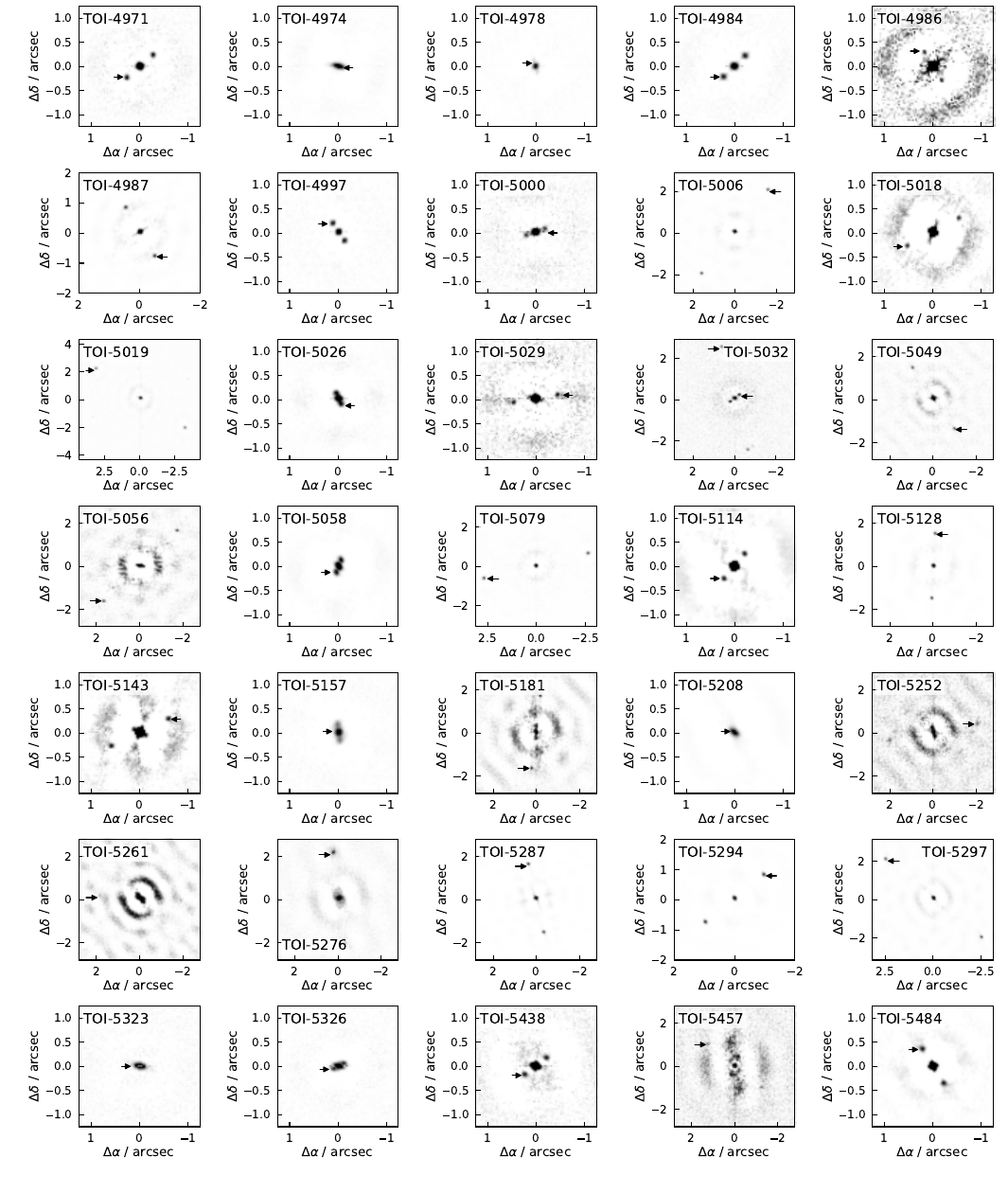}
\caption{Continuation of Figure~\ref{fig:newacf10} for TOI-4971 through TOI-5484.\label{fig:newacf11}}
\end{figure*}

\begin{figure*}[p]
\centering
\includegraphics[width=0.93\textwidth]{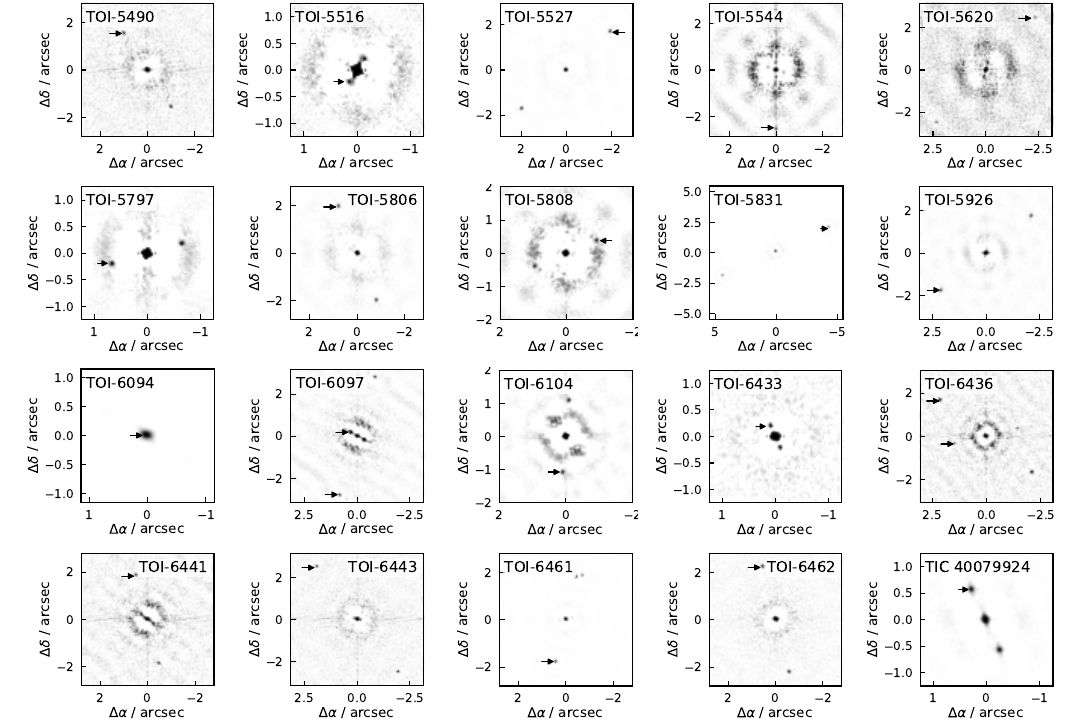}
\caption{Continuation of Figure~\ref{fig:newacf11} for TOI-5490 through TOI-6462 and the unnumbered TOI host TIC~40079924.\label{fig:newacf12}}
\end{figure*}

\clearpage
\subsection{Gaia DR3 Common-Motion Companions}\label{app:gaia}

Table~\ref{tab:gaiacomplete} shows representative rows from the 510 companions selected by the uniform Gaia DR3 common-motion search. The complete table is supplied in machine-readable form and includes both Gaia source identifiers and all quantities used in the selection; the compact printed version gives the quantities most directly used in the combined census.

\startlongtable
\input{sample_gaia_dr3_companions.tex}

\clearpage
\twocolumngrid
\bibliographystyle{aasjournal}
\bibliography{references}

\end{document}

%% file: appendix_context.tex
\section{Selection and Catalog Context}\label{app:selectionfigures}
\suppressfloats[t]

\begin{figure}[h!]
\centering
\includegraphics[width=0.94\textwidth]{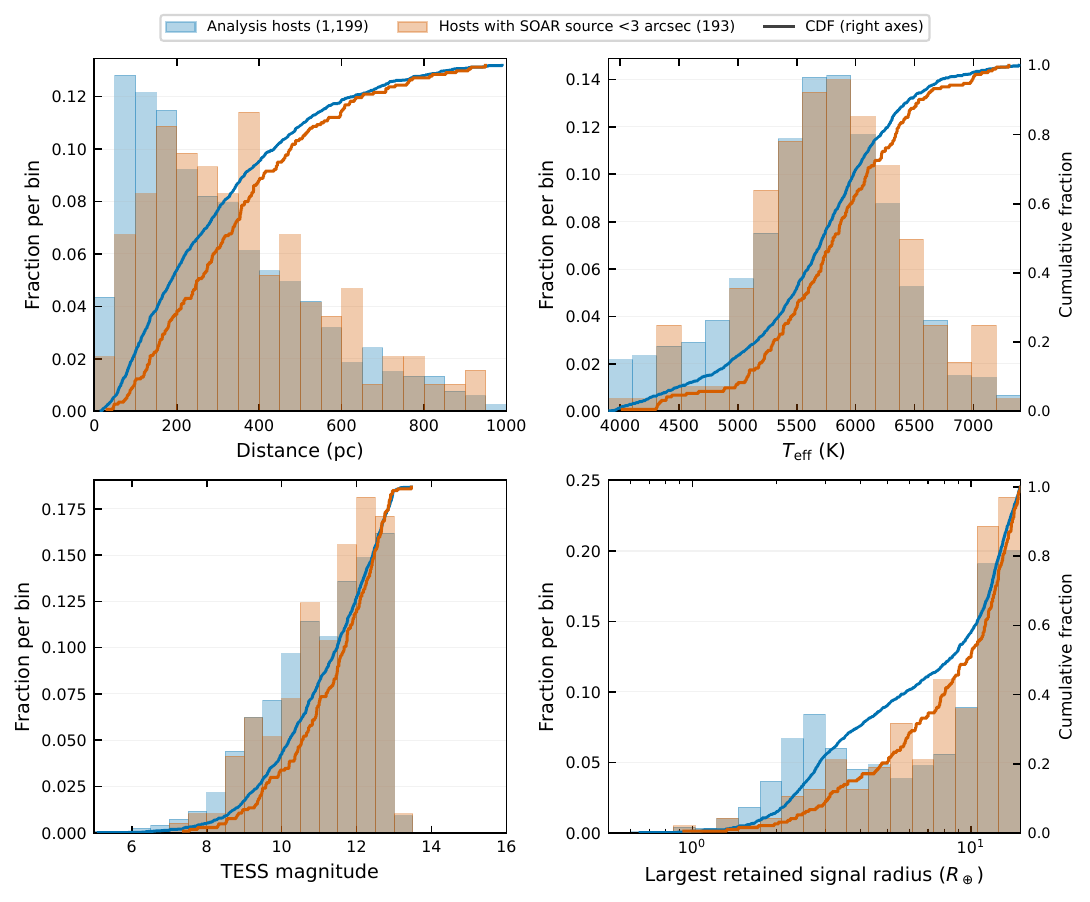}
\caption{Distributions of distance, effective temperature, TESS magnitude, and the largest retained catalog planet or candidate radius for the cumulative 1,199-host mass-ratio-matched demographic sample and its 193-host subset with a resolved SOAR source inside 3 arcsec. Translucent bars give the normalized fraction in each bin on the left axes; solid step curves give the empirical cumulative distributions on the right axes. The resolved-source subset is shifted toward larger distances and larger catalog candidate radii. Raw imaging yields are therefore not occurrence rates.}\label{fig:properties}
\end{figure}

Within the cumulative 1,199-host dwarf-primary demographic sample, 193 hosts have at least one resolved SOAR source inside 3 arcsec. This resolved-source subset is, on average, more distant and more likely to contain a large-radius candidate than the full demographic sample (Figure~\ref{fig:properties}). Binned distributions show where each shift is concentrated, while the overlaid empirical cumulative distributions show the integrated difference without relying on a particular bin edge. They reflect heterogeneous TOI selection and residual false-positive contamination and motivate the target-specific sensitivity model used in the main analysis.

\begin{figure}[h!]
\centering
\includegraphics[width=0.94\textwidth]{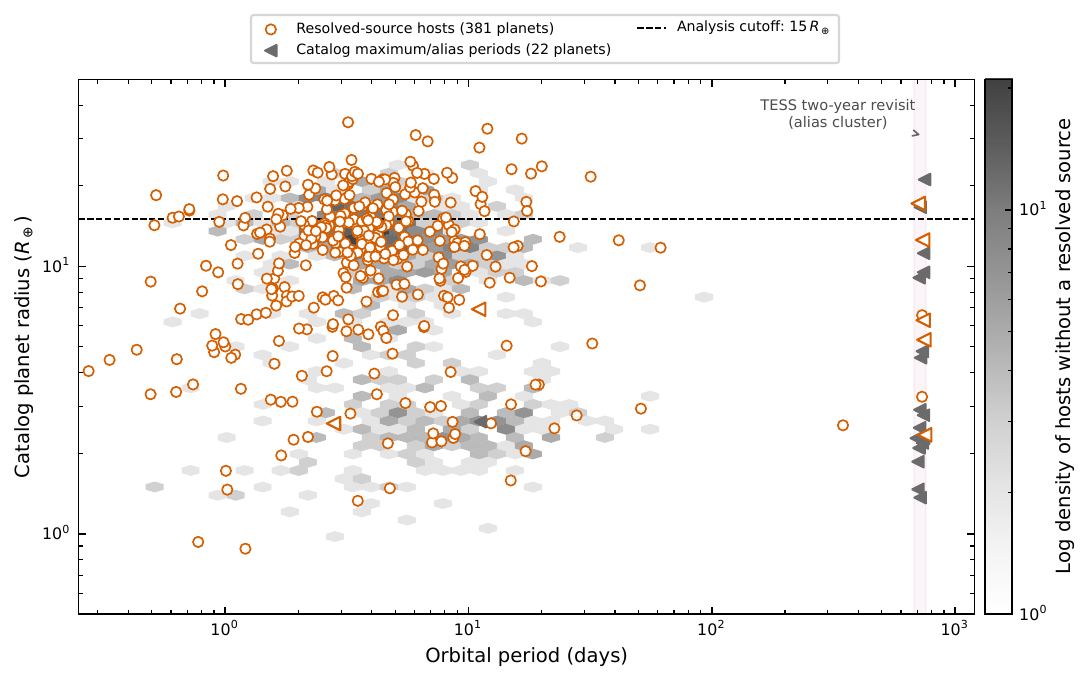}
\caption{Catalog orbital period and radius for current CP, KP, PC, and APC entries associated with cumulative SOAR targets. The hexagonal-bin shading shows the full catalog density, and orange points mark entries whose host has a resolved source inside 3 arcsec. The dashed line is the 15~\rearth\ cutoff used for the close-binary sample. Left-pointing triangles mark the 22 entries whose catalog notes identify their 680--760 day values as maximum or integer-alias periods; the shaded band marks the TESS Year-1-to-Year-3 revisit window. Multiple planets in one system are plotted separately.}\label{fig:periodradius}
\end{figure}

The frozen planet table contains 2,351 retained CP, KP, PC, or APC entries around 2,197 surveyed TICs, including 377 confirmed-planet entries around 312 TICs. Figure~\ref{fig:periodradius} shows their descriptive period--radius distribution. Companion hosts occur throughout the diagram, but large candidates are overrepresented because suspected false positives were preferentially imaged. The apparent column near 700--750 days arises from the TESS Year-1-to-Year-3 revisit window: 22 of the 25 catalog entries between 680 and 760 days explicitly state that the listed value is a maximum or integer-alias period inferred from sparse transits. Their actual periods lie to the left of the plotted values, often at an integer divisor. We retain the frozen catalog entries but mark those 22 aliases separately. The figure provides catalog context and is excluded from the occurrence inference.

%% file: appendix_secondary.tex
\clearpage
\section{Secondary and Exploratory Analyses}\label{app:secondary}
\suppressfloats[t]

\subsection{Mass-ratio dependence}\label{sec:massratio}
\suppressfloats[t]

The main demographic sample is restricted to the 1,199 dwarf primaries for which the empirical mass--luminosity conversion is applicable. For each host, we interpolate the 2022 April 16 version of the dwarf sequence from \citet{PecautMamajek2013}, using $M_I=M_V-(V-I_C)$, to convert measured $\Delta I$ to a candidate-secondary mass.\footnote{\href{https://www.pas.rochester.edu/~emamajek/EEM_dwarf_UBVIJHK_colors_Teff.txt}{Pecaut--Mamajek dwarf table}, version 2022 April 16, accessed 2026 July 31.} The same relation sets every simulated contrast, HRCam detection, and magnitude-limited weight. Across these primaries, $q=0.4$ corresponds to $\Delta I=4.25$ mag, with a 16th--84th percentile range of 3.88--4.41 mag. The primary result applies this target-specific mapping directly.

Forty-eight empirical $q\geq0.4$ companions remain inside 100 au. The fiducial field calculation uses the same \citet{ElBadryRix2019} primary-mass- and projected-separation-dependent law as the main response. Thirteen companions are observed at $0.4\leq q<0.7$ compared with 28.35 expected, while 35 are observed at $0.7\leq q\leq1$ compared with 96.10 expected. Their relative frequencies are 0.482 (0.364--0.623) and 0.371 (0.313--0.436), respectively. Conditional on 48 total detections and this field distribution, the probability of obtaining 35 or fewer high-$q$ companions is 0.288. Together with the post hoc bin boundary, this result supplies no evidence for different suppression in the two broad $q$ intervals.

\begin{figure}[t]
\centering
\includegraphics[width=0.98\textwidth]{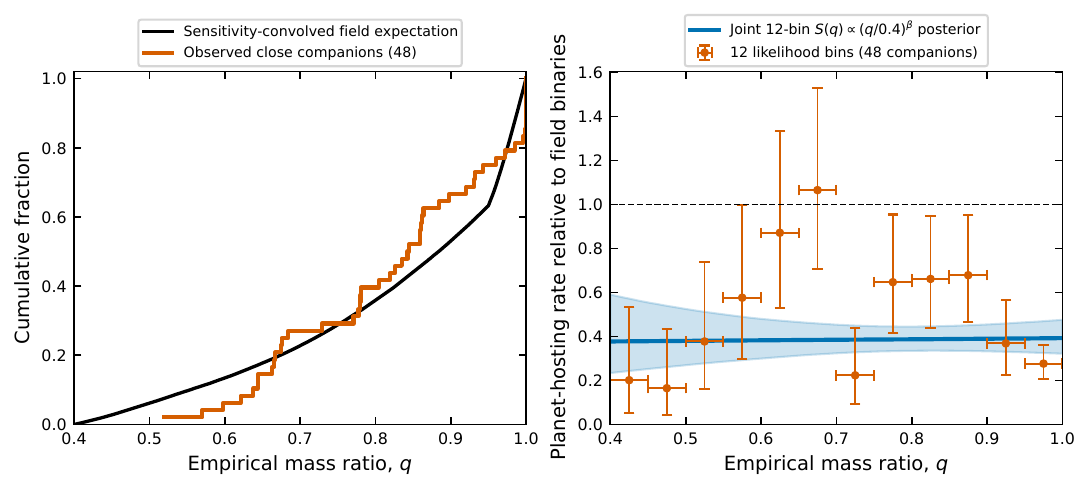}
\caption{Exploratory mass-ratio analysis for dwarf primaries and projected separations below 100 au. Left: empirical cumulative $q$ distribution of the 48 detected companions and the target-contrast-convolved El-Badry field expectation. Right: posterior for the continuous law in Equation~\ref{eq:qfit}; orange points show the exact 12 equal-width bin counts used jointly in the conditional likelihood, with 68\% Poisson intervals. The blue band gives uncertainty in the pooled mean law; individual-bin prediction intervals would be wider. Alternate bin counts give consistent uncertainty, while simplified field-$q$ laws can change the fitted sign. The present data remain consistent with mass-ratio-independent suppression.}\label{fig:massratio}
\end{figure}

Our continuous shape test fits
\begin{equation}
S(q)=C\left(\frac{q}{0.4}\right)^\beta
\label{eq:qfit}
\end{equation}
to 12 equal-width mass-ratio bins, conditioning the slope likelihood on the 48 observed companions. The likelihood is evaluated directly from the 12 integer bin counts shown in Figure~\ref{fig:massratio}; the two broad $q$ intervals quoted above serve only as descriptive summaries. Repeating the calculation with 6, 8, 10, 15, 20, and 24 equal-width bins moves the posterior median only from 0.001 to 0.165, and every 95\% interval includes zero. The fiducial 12-bin posterior is $\beta=0.04^{+0.65}_{-0.61}$, with a 95\% interval of $-1.14<\beta<1.36$ and $P(\beta<0)=0.469$. A likelihood-ratio test against $\beta=0$ gives $p=0.991$; the test compares the best-fitting free slope with the zero-slope model and asks whether the improvement in likelihood is larger than expected from adding one fitted parameter. The results are fully consistent with a mass-ratio-independent suppression amplitude.

The field $q$ distribution remains the dominant interpretive uncertainty. As sensitivity tests, we replaced the El-Badry law with simple power-law indices $-0.5\leq\gamma\leq0.5$ and twin fractions of 0--20\%, spanning commonly used solar-type approximations \citep{Raghavan2010,MoeDiStefano2017}. The fitted $\beta$ then ranges from $-1.11$ to $+1.53$, and the likelihood-ratio probability ranges from 0.030 to 0.740 (Table~\ref{tab:qprior}). Both the sign and nominal significance depend on the assumed field population. Multi-band contrasts and target-specific ages and metallicities are needed before a physical $q$ dependence can be tested more sharply.

\begin{deluxetable*}{lrrrrr}
\tablecaption{Field mass-ratio model sensitivity\label{tab:qprior}}
\tablehead{\colhead{Field $q$ law}&\colhead{$\gamma$}&\colhead{$f_{\rm twin}$}&\colhead{$\beta_{50}$}&\colhead{95\% interval}&\colhead{$p_{\rm LRT}$}}
\startdata
El-Badry $p(q\mid M_1,s)$&\nodata&\nodata&$0.04$&$-1.14,\ 1.36$&0.991\\
Simple power law&$-0.5$&0.00&$1.53$&$0.17,\ 3.08$&0.030\\
Simple power law&$-0.5$&0.10&$-0.17$&$-1.32,\ 1.11$&0.740\\
Simple power law&$0.0$&0.00&$1.03$&$-0.33,\ 2.58$&0.159\\
Simple power law&$0.0$&0.10&$-0.29$&$-1.47,\ 1.05$&0.621\\
Simple power law&$0.5$&0.10&$-0.52$&$-1.74,\ 0.85$&0.406\\
Simple power law&$0.0$&0.20&$-1.11$&$-2.22,\ 0.12$&0.067\\
\enddata
\tablecomments{The first row is the fiducial mass- and separation-dependent field law. For the alternatives, the density is proportional to $q^\gamma$ plus a uniform excess at $q>0.95$ containing fraction $f_{\rm twin}$. The last column is the likelihood-ratio-test probability for $\beta=0$. The simple laws are sensitivity tests, not equally weighted posterior models.}
\end{deluxetable*}

The mass-ratio analysis supports the close-source deficit itself, while any variation of its amplitude with companion mass ratio remains unresolved.

\clearpage

%% file: appendix_model_fits.tex
\section{Numerical Model-fitting Details}\label{app:fitdetails}
\suppressfloats[t]

Reproducing the separation-recovery results requires two distinct numerical calculations. The HRCam-only analysis estimates a posterior distribution for the smooth-law parameters, whereas the combined SOAR--Gaia analysis finds maximum-likelihood parameters for model comparison. The procedures below make those related statistical outputs explicit.

\subsection{HRCam-only recovery fit}

For each trial value of $\boldsymbol{\theta}=(S_0,a_{50},w)$, Equation~\ref{eq:recovery} weights the simulated field population in physical semimajor axis. Let $R_{jm}$ be the corrected simulated field contribution from semimajor-axis bin $m$ to projected-separation bin $j$ after orbital projection and target contrast selection. Before $R_{jm}$ is accumulated, every synthetic pair receives the truncated-Gaussian eccentricity weight and, from 50 to 5,000 au, its own El-Badry mass- and projected-separation-dependent $q$ weight. A pair above the angular/contrast locus in Equation~\ref{eq:gaiaangular} also has its own combined-light volume factor divided out as specified by Equation~\ref{eq:widesel}. The expected count is
\begin{equation}
\lambda_j(\boldsymbol{\theta})=\sum_m R_{jm}S(a_m;\boldsymbol{\theta}).
\label{eq:projectedexpectation}
\end{equation}
The step, smooth, and constant models use the same corrected response matrix. Suppression is therefore applied in physical semimajor axis, while the field-population and resolved-target weights are fixed event by event before projected-separation binning. Each model is then tested against the same observed projected separations.

For observed count $n_j$ in projected bin $j$, the independent-bin Poisson log-likelihood is
\begin{equation}
\ln \mathcal{L}(\boldsymbol{\theta})=
\sum_j\left[n_j\ln\lambda_j(\boldsymbol{\theta})-
\lambda_j(\boldsymbol{\theta})-\ln(n_j!)\right].
\label{eq:poissonbins}
\end{equation}
This likelihood is appropriate because the forward model predicts the mean count in each disjoint bin. Orbital projection and survey completeness enter through $\lambda_j$.

We adopted uniform priors in $S_0$ and $\log_{10}a_{50}$ over $0.001<S_0<0.5$ and $3.2<a_{50}<1000$ au, and a log-uniform prior over $0.03<w<1$ dex. The logarithmic priors assign equal weight to equal multiplicative ranges. These bounds span nearly zero to moderate inner suppression and transitions from nearly step-like to broad. The posterior remains well below the upper $S_0$ boundary, and all reported intervals are conditional on the priors and fixed outer asymptote.

We evaluated 81 linearly spaced values of $S_0$, 101 linearly spaced values of $\log_{10}(a_{50}/{\rm au})$, and 51 logarithmically spaced values of $w$, for $417{,}231$ parameter combinations. Equal grid-cell weights implement the stated priors. At each point we applied Equations~\ref{eq:recovery}--\ref{eq:poissonbins}, normalized $\exp(\ln\mathcal{L}-\ln\mathcal{L}_{\max})$, and marginalized by summing over the other grid dimensions. The reported values are posterior medians with 16th--84th percentile intervals. The predictive band in Figure~\ref{fig:continuous} uses 8,000 draws from the normalized discrete posterior with seed 260731; the occurrence calculation uses 50,000 draws from the same posterior. These draws propagate parameter uncertainty while holding the observed Poisson counts fixed.

For the AIC comparison, the fitted step model used 151 inner multipliers from 0.001 to 0.60 and 251 values of $\log_{10}(a_{\rm cut}/{\rm au})$ from 0.5 to 3.0. The constant model used 1,200 multipliers from 0.001 to 1.20. AIC was calculated from the highest Poisson likelihood on each grid. The Paper II step shown in Figure~\ref{fig:continuous} is plotted only as a literature benchmark and was not included among the fitted models.

\subsection{Combined SOAR--Gaia model comparison}

The combined analysis uses the same 0.25 dex projected-separation bins and Poisson likelihood over 1--5,000 au. Each trial binary distribution is passed through the appropriate SOAR and Gaia selection matrices, with the same event-level eccentricity, El-Badry mass-ratio, and angular resolved-target weights, before it is evaluated with the likelihood. These externally specified weights have no fitted parameters, so they contribute nothing to $k_{\rm par}$ in AIC. Unlike the HRCam-only posterior grid, this analysis estimates maximum-likelihood parameters for AIC.

We minimized the negative Poisson log-likelihood with SciPy's \texttt{differential\_evolution} global optimizer, using seed 120812, tolerance $10^{-9}$, and local polishing. A global optimizer was chosen because the scale, width, and amplitude parameters can be correlated. The fixed seed makes the stochastic search repeatable, while local polishing refines the best global solution.

The smooth-model bounds were $0.001\leq S_0\leq1$, $0\leq\log_{10}(a_{50}/{\rm au})\leq3.5$, and $0.03\leq w\leq1.5$ dex; the step model used the same amplitude and scale bounds. Shifted models allowed the field mean $\log_{10}P$ to vary from 5.03 to 7.0 and either fixed its 2.28 dex width or allowed widths down to 0.30 dex. The 100 au benchmark assumes $1.6\,\msun$ total mass. Its hybrid version fits amplitude over 0.05--1.5 and width over 0.30--2.28 dex. Only this hybrid can reduce the total multiplicity; the pure shifted models conserve it.

For each model we calculated ${\rm AIC}=2k_{\rm par}-2\ln\mathcal{L}_{\max}$, where $k_{\rm par}$ counts only parameters fitted to the combined data. These fitted quantities are maximum-likelihood values, not posterior medians or credible intervals. They should therefore not replace the HRCam-only posterior values reported in Section~\ref{sec:continuous}.

%% file: sample_new_companion_measurements.tex
\begin{deluxetable*}{rrllrrrr}
\tabletypesize{\fontsize{6.3}{7.0}\selectfont}
\tablewidth{0pt}
\tablecaption{Representative SOAR/HRCam companion measurements\label{tab:newmeasurements}}
\tablehead{
\colhead{TIC} & \colhead{TOI} & \colhead{UT date} & \colhead{Pair} &
\colhead{$\rho$} & \colhead{P.A.} & \colhead{$\Delta I$} & \colhead{$P_{\rm bg}$} \\
\colhead{} & \colhead{} & \colhead{} & \colhead{} &
\colhead{(arcsec)} & \colhead{(deg)} & \colhead{(mag)} & \colhead{}
}
\startdata
235037761 & 131  & 2022 May 22 & A--B   & 0.1431 & 17.3  & 1.2 & 0\\
220396259 & 379  & 2021 Oct 01 & A--B   & 0.0602 & 18.7  & 0.3 & 0\\
176984236 & 405  & 2021 Nov 20 & A--B   & 0.3865 & 321.7 & 0.9 & 0\\
69068526  & 1196 & 2021 Apr 25 & A--BC  & 1.6777 & 198.5 & 2.4 & 0\\
69068526  & 1196 & 2021 Apr 25 & B--C   & 0.2203 & 210.7 & 1.8 & \nodata\\
201508515 & 2539 & 2021 Oct 01 & A--C   & 1.6169 & 272.1 & 4.1 & 0\\
191496430 & 2909 & 2021 Nov 20 & Aa--Ab & 0.1842 & 234.9 & 2.4 & 0\\
448358913 & 3496 & 2021 Nov 20 & Ba--Bb & 0.4030 & 153.0 & 2.3 & \nodata\\
31637479  & 4182 & 2021 Nov 20 & A--B   & 0.0371 & 235.8 & 0.2 & 0\\
31637479  & 4182 & 2024 Jan 08 & A--B   & 0.0452 & 223.2 & 0.0 & 0\\
308504559 & 5831 & 2023 Aug 31 & A--B   & 4.7723 & 294.5 & 0.7 & 0\\
35883621  & 6094 & 2024 Jan 08 & A--B   & 0.0378 & 81.2  & 0.7 & 0\\
123133954 & 6461 & 2024 Jan 08 & A--BC  & 1.8235 & 346.3 & 1.3 & 0\\
123133954 & 6461 & 2024 Jan 08 & B--C   & 0.2616 & 281.1 & 0.5 & \nodata\\
\enddata
\tablecomments{The complete 433-row table is available in machine-readable form. A is the catalog primary; A--B and A--C measure B and C relative to A; Aa--Ab is an inner pair within A; A--BC measures A relative to an unresolved BC subsystem; and B--C or Ba--Bb measure an inner pair within the B subsystem. $P_{\rm bg}$ is the local Gaia chance-alignment probability defined by Equation~\ref{eq:pbg}. Ellipses indicate that $P_{\rm bg}$ is unavailable because the row is an inner-subsystem measurement, required Gaia information is missing, or the available comparison cone does not extend beyond 10 arcsec. The machine-readable table also gives distance, projected separation, filter, $X_{R,A}$, local-density inputs, and position-angle provenance. Astrometric precision is retained from the nightly HRCam reductions.}
\end{deluxetable*}

%% file: sample_gaia_dr3_companions.tex
\begin{deluxetable*}{rrrrr}
\tabletypesize{\scriptsize}
\tablewidth{0pt}
\tablecaption{Representative Gaia DR3 common-motion companions\label{tab:gaiacomplete}}
\tablehead{
\colhead{TIC} & \colhead{$\rho$} & \colhead{$s_{\rm proj}$} &
\colhead{$\Delta G$} & \colhead{$\gamma_{\rm CPM}$} \\
\colhead{} & \colhead{(arcsec)} & \colhead{(au)} & \colhead{(mag)} & \colhead{}
}
\startdata
453260209 & 1.76 & 60.9  & 0.12  & 29.9\\
364107753 & 1.33 & 68.9  & 4.73  & 29.8\\
952046774 & 1.61 & 75.0  & 0.18  & 38.0\\
322063810 & 2.64 & 81.5  & 4.20  & 48.3\\
29919341  & 0.85 & 85.7  & 0.87  & 42.8\\
280095254 & 0.83 & 86.2  & 1.54  & 198.0\\
406672232 & 2.68 & 124.8 & 3.46  & 10.6\\
263003176 & 2.25 & 129.5 & 5.81  & 32.6\\
409934330 & 0.80 & 139.3 & 1.06  & 9.6\\
166184426 & 6.30 & 141.6 & $-0.19$ & 93.8\\
391949880 & 2.20 & 150.7 & 2.58  & 52.3\\
133334108 & 2.39 & 154.4 & 5.24  & 44.2\\
\enddata
\tablecomments{The complete 510-row table is available in machine-readable form. It includes both Gaia DR3 source identifiers, astrometry, photometry, uncertainties, and all common-motion selection diagnostics.}
\end{deluxetable*}